# Unifying concepts in information-theoretic time-series analysis

**Annie G. Bryant** [1,2,✉], **Oliver M. Cliff** [1,2], **James M. Shine** [2,3], **Ben D. Fulcher** [1,2], **and Joseph T. Lizier** [2,4]

[1]School of Physics, The University of Sydney, Camperdown, NSW, Australia
[2]Centre for Complex Systems, The University of Sydney, Camperdown, NSW, Australia
[3]Brain and Mind Centre, The University of Sydney, Camperdown, NSW, Australia
[4]School of Computer Science, The University of Sydney, Camperdown, NSW, Australia

Correspondence: *annie.bryant@sydney.edu.au*

## Abstract

Information theory is a powerful framework for quantifying complexity, uncertainty, and dynamical structure in time-series data, with widespread applicability across disciplines such as physics, neuroscience, and finance. However, the literature on these measures remains fragmented, with domain-specific terminologies, inconsistent mathematical notation, and disparate visualization conventions that hinder interdisciplinary integration. This work addresses these challenges by unifying key information-theoretic time-series measures through shared semantic definitions, standardized mathematical notation, and cohesive visual representations. We compare these measures in terms of their theoretical foundations, computational formulations, and practical interpretability—mapping them onto a common conceptual space through an illustrative case study with functional magnetic resonance imaging time series in the brain. This case study exemplifies the complementary insights these measures offer in characterizing the dynamics of complex neural systems, such as signal complexity and information flow. By providing a structured synthesis, our work aims to enhance interdisciplinary dialogue and methodological adoption, which is particularly critical for accessibility and reproducibility in computational neuroscience. More broadly, our framework serves as a resource for researchers seeking to navigate and apply information-theoretic time-series measures to diverse complex systems.

## Code accompaniment

All code used for empirical analysis and visualization in this work is freely available at `https://github.com/anniegbryant/info_theory_visuals`.

## The need for a cohesive, accessible summary of information-theoretic time-series measures

Information theory, originally developed for studying communications channels, provides a model-free statistical framework[1] for analyzing variance within and relationships between random variables [1–3]. Measures derived from information theory, like mutual information and transfer entropy, present powerful and domain-general tools with which researchers can quantify distinctive patterns of information flow and dependence, with particular utility in applications to multivariate time series samples of random variables. Information-theoretic measures are increasingly developed and applied across diverse domains, from economics [4] to epidemiology [5] to neuroscience [6, 7]. Despite this smorgasbord of different measures at a researcher's disposal—including different forms of entropy, directed information flow, and stochastic interactions—we are generally limited to a disjoint literature that employs inconsistent terminology and mathematical notations. These differences in nomenclature create a sizable learning curve for a researcher to understand the nuances of a single measure, which becomes even steeper when trying to understand intricate similarities and differences across multiple measures. As a result, one may select a measure based on subjective criteria and/or domain-specific conventions, potentially missing more appropriate information-theoretic measure(s) for the problem at hand. Prior work has endeavored to provide practical guides for applying and interpreting general information-theoretic measures (including joint/conditional entropy, mutual information, transfer entropy, and partial information decomposition) across disciplinary domains [7–12], but a unifying treatment of how information-theoretic methods can be adapted to time-series analysis remains lacking.

To our knowledge, no prior work has attempted to unify the disjoint literature on information-theoretic measures for time series as a coherent presentation using consistent terminology and visual elements. To address this gap, here we conceptually unify information-theoretic time-series measures with common semantic nomenclature and mathematical notation in a cohesive guide that offers accessible categorization and explanations. We set two aims with this paper: first, we aim to unify information-theoretic measures into a common schematic and tabular representation to provide a high-level summary that is broadly accessible to interdisciplinary readers interested in applying information theory to time series data. Second, we aim to guide the reader interested in more in-depth coverage of a particular measure through the computation and interpretation of that measure when applied to empirical time series data (exemplified with functional neuroimaging data), bridging accessible non-technical ideas with the mathematical expressions that encapsulate those ideas. As such, our approach is to actively incorporate the mathematical formulations of each measure alongside plain-language summaries and interpretable visuals to improve the accessibility of these powerful measures across disciplines. Visual schematics form the basis for much of this paper, motivated by work demonstrating how visual explanations facilitate conceptual links between the structural components that collectively interact to form a complex time-varying system [13].

## Getting down to brass tacks...onomy

Before delving into the information-theoretic measure taxonomy, it will be helpful to first outline a few key definitions. We define a 'time series' as a time-indexed stochastic process, comprised of a sequence of random variables $\mathtt{X}$ = $\{X_1, X_2, X_3, \ldots, X_T\}$ (where $T$ is the length of the process) as depicted schematically in Fig. 1. Each $X_t$ is a random variable that can take the form of a single dimension (univariate), two dimensions (bivariate), or multiple dimensions (multivariate).[2] Here, we primarily consider the case where each process ($\mathtt{X}$) is univariate for simplicity, though any of the variables in any of the measures could be trivially made bivariate or more generally multivariate.

In the context of brain dynamics, such a general process might refer abstractly to time-series measurement of the neural activity in a brain region, evolving over time. We can take repeated measurements of, or in other words, sample this process, which corresponds to the 'Applied' angle in Fig. 1. The sample path, known as a 'realization', of the time series is denoted $\mathtt{x}$ and comprises an ordered collection of individual empirical measurements $\{x_1, x_2, x_3, \ldots, x_T\}$. Within this empirically measured time series (realization), each $x_t \in X_t$ is a realization of the corresponding random variable. In other words, the empirical measurement at $x_1$ is a specific realization of the random variable $X_1$.

[1] 'Model-free' here, as generally phrased in the literature, is used to mean that information-theoretic measures are more flexible across a range of potential probability distributions in less parameter-dependent ways than other methods. While stronger assumption-based models (like the Pearson correlation coefficient) constrain the probabilistic forms that can be captured, information-theoretic measures do not constrain the form of the relationship; instead, their estimators can (in principle) adapt to the nature of any relationship. As we shall see later in the text, advanced estimators do this with weaker parametric assumptions, and so we switch our phrasing to 'model-general'.

[2] In the multivariate case, for a process of dimension $m$, we would use vector notation for the random variable at time $t$: $\mathbf{X}_t = \{X_{t,1}, X_{t,2}, \ldots, X_{t,m}\}$.

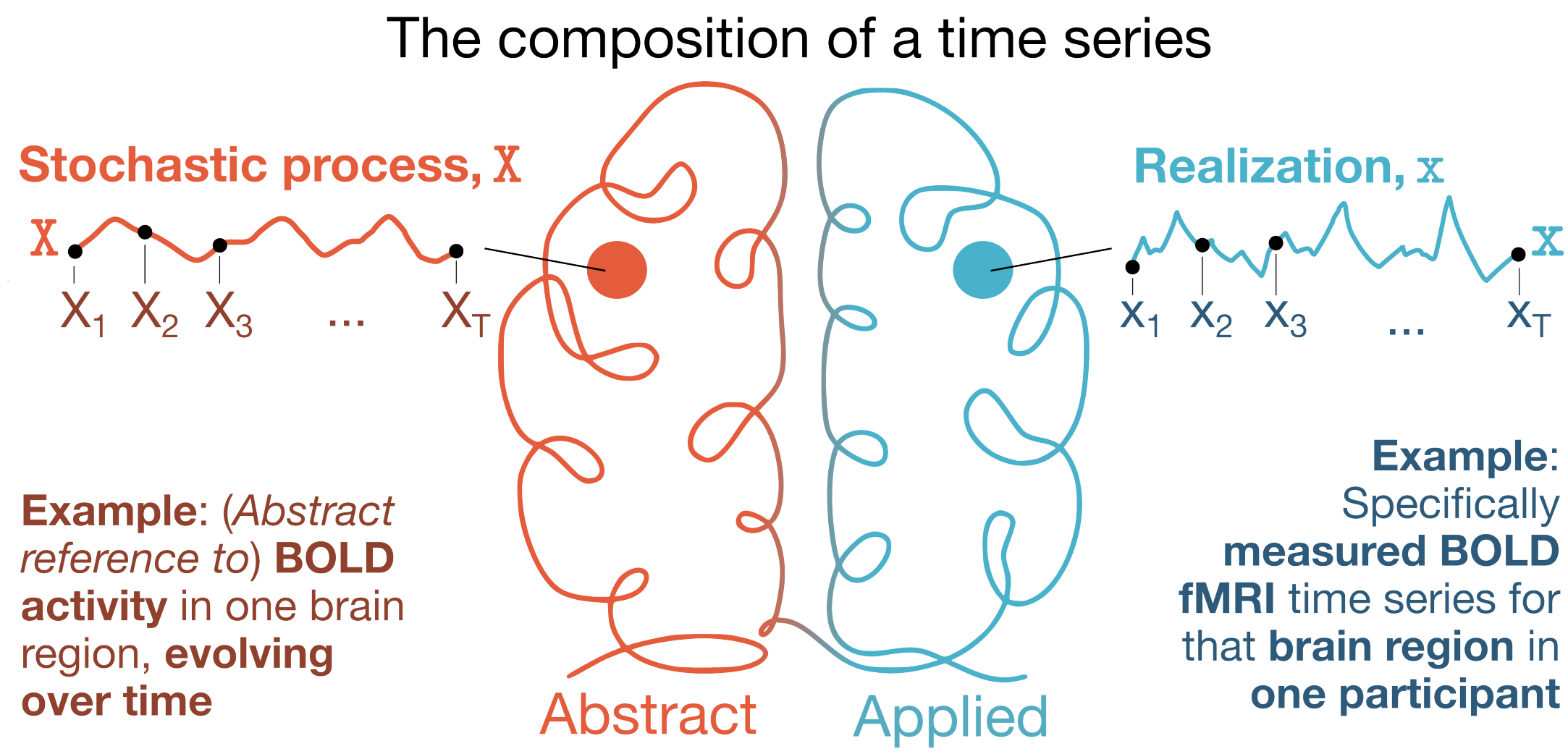

**Figure 1. Decomposing the elements of a time series, in the context of a stochastic process and a specific realization.** Here, we start from the more abstract concept of a stochastic process, $\mathtt{X}$, which is an ordered collection of random variables $\{X_1, X_2, X_3, \ldots, X_T\}$ where $T$ is the total length of the process. For example, in the context of brain dynamics, the stochastic process $\mathtt{X}$ might refer abstractly to the measurement of neural activity in a brain region, evolving over time. In parallel, on the applied end, a particular observation of our stochastic process is known as a realization, denoted $\mathtt{x}$. This realization is an ordered collection of individual empirical measurements $\{x_1, x_2, x_3, \ldots, x_T\}$. Continuing the above example, the realization $\mathtt{x}$ might reflect the measured blood oxygen level-dependent (BOLD) functional magnetic resonance imaging (fMRI) time series in the given brain region in one participant.

This distinction between the underlying stochastic process $\mathtt{X}$ and an individual empirical realization $\mathtt{x}$ is important, particularly when interpreting measured time series data. In most cases—for example, in human functional neuroimaging—we typically only have access to a single realization of a process, and analyses therefore often proceed under a working assumption of ergodicity. Here, 'ergodic' means that the realization ($\mathtt{x}$) will replicate the probabilities of the space of the process ($\mathtt{X}$). In other words, over sufficient time, the time-averaged behavior of realization $\mathtt{x}$ will converge to the true ensemble-averaged behavior of the underlying process $\mathtt{X}$ [14]. Under this assumption of ergodicity, the realizations $x_1, x_2, x_3, \ldots, x_T$ (of what would otherwise be distinct random variables) may be treated as multiple realizations $x$ of a single, time-independent random variable $X$ associated with process $\mathtt{X}$. This approximation enables the application of information-theoretic measures to empirically observed time series by allowing averages computed over a single realization to serve as practical estimates of ensemble averages over the process distribution. We note, however, that for finite samples this may require very long time series to adequately capture the probabilistic properties of the process—a condition that is not always satisfied by, for example, short neuroimaging time series measured at low temporal resolution, and which therefore warrants additional care. For a more thorough discussion of this ergodicity assumption and relevant reading, see the Methods section 'Advanced concepts in estimation'.

We define a 'discrete' random variable as one that exhibits a countable number of distinct values, while a 'continuous' random variable exhibits an uncountable number of possible values. Each outcome within the finite set for a discrete variable is associated with a probability mass, and a probability density function (PDF) links probabilities with these finite outcomes. We introduce and interpret the mathematical notation for each measure throughout this piece, which is generally predicated on the probability mass $p(x_t = a_i)$ for a discrete variable. This distribution captures the probability of obtaining the $i$-th value $x_t = a_i$ at time $t$ across all time points $\{1 \ldots T\}$. For a continuous variable, the probability mass of obtaining a value $b > x_t > a$ is an integral over the probability density function $f(x)$: $p(b > x_t > a) = \int_a^b f(x)dx$. We assume that the PDF is defined for the time series variable at hand—or variables, where the measure is multivariate. A maximum likelihood estimator (MLE) can be used to estimate the probability mass from observed empirical data for discrete variables.[3] As the information-theoretic measures we consider are generally constructed from sums and differences of entropies, their MLE follows directly from the PDF. We primarily consider discrete random values in conveying the conceptual properties of information-theoretic time-series measures, though we extend to continuous in our illustrative case study with resting-state brain activity in each section of this work.

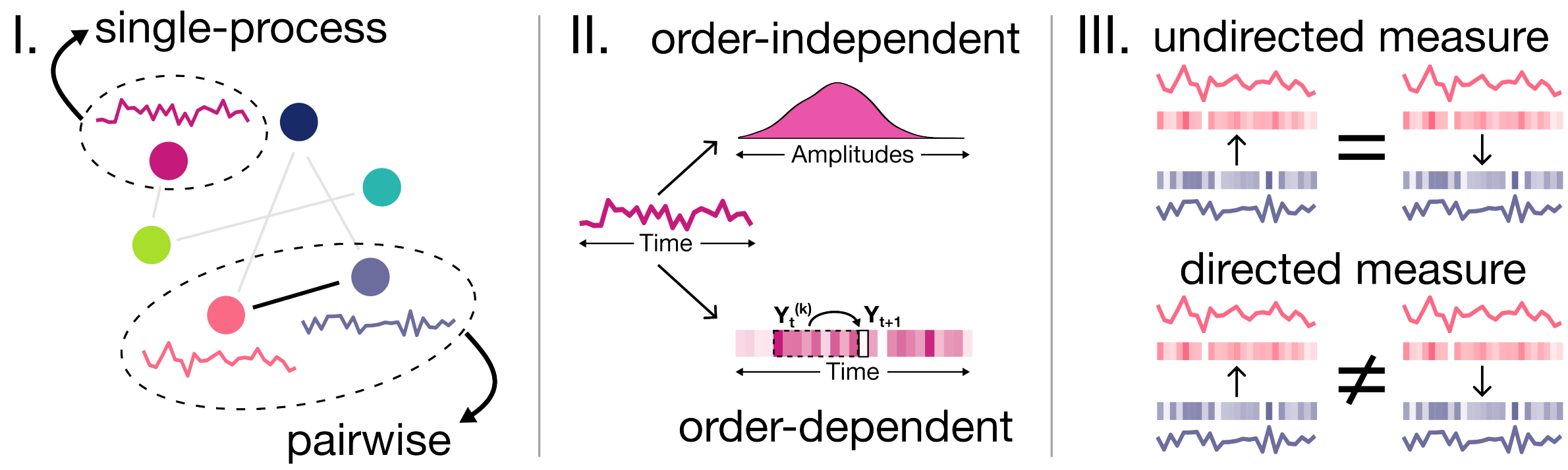

**Figure 2. Three key characteristics distinguish the information-theoretic time-series measures covered here.** Each panel illustrates a conceptual characteristic that separates classes of information-theoretic time series measures covered in this review. **I.** Single-process versus pairwise: whether the measure describes properties of a single time series (single-process) or relationships between two time series (pairwise). **II.** Order-independent versus order-dependent: whether temporal ordering is taken into account, distinguishing measures based on static distributions (order-independent) from those sensitive to dynamics and/or time-lagged structure (order-dependent). **III.** Undirected versus directed: for the case of pairwise measures specifically, whether the measure captures symmetric relationships (undirected) or directional influence (directed).

With these definitions in hand, we can categorize the existing literature of information-theoretic time-series measures into a sensible and interpretable 'taxonomy'. Our taxonomy is organized based on three key characteristics, as schematically depicted in Fig. 2 and detailed below:

I. Single-process versus pairwise: A single-process measure operates on a single process (e.g., the activity of a single brain region) while a pairwise measure captures dependencies between two processes (e.g., the activity in two brain regions). Higher-order (three or more variable) information measures, including interaction information [2, 15] and partial information decomposition frameworks [16], are beyond the scope of the present taxonomy; readers are referred to those references for comprehensive treatments.

II. Order-independent versus order-dependent: An order-independent measure operates solely on the distribution of time series values without considering the order in which the underlying values occurred. By contrast, an order-dependent measure is sensitive to the temporal order in which the values occurred.

III. Undirected versus directed (pairwise measures only): An undirected measure captures equivalent dependence (i.e., symmetric) between processes $X$ and $Y$. By contrast, a directed measure is computed specifically from $X$ to $Y$, such that the value from $Y$ to $X$ is not necessarily symmetric.

[3] More sophisticated estimators exist (e.g., based on a Bayesian framework) but remain out of scope for this review.

## An all-in-one schematic and tabular reference for key information-theoretic time-series measures

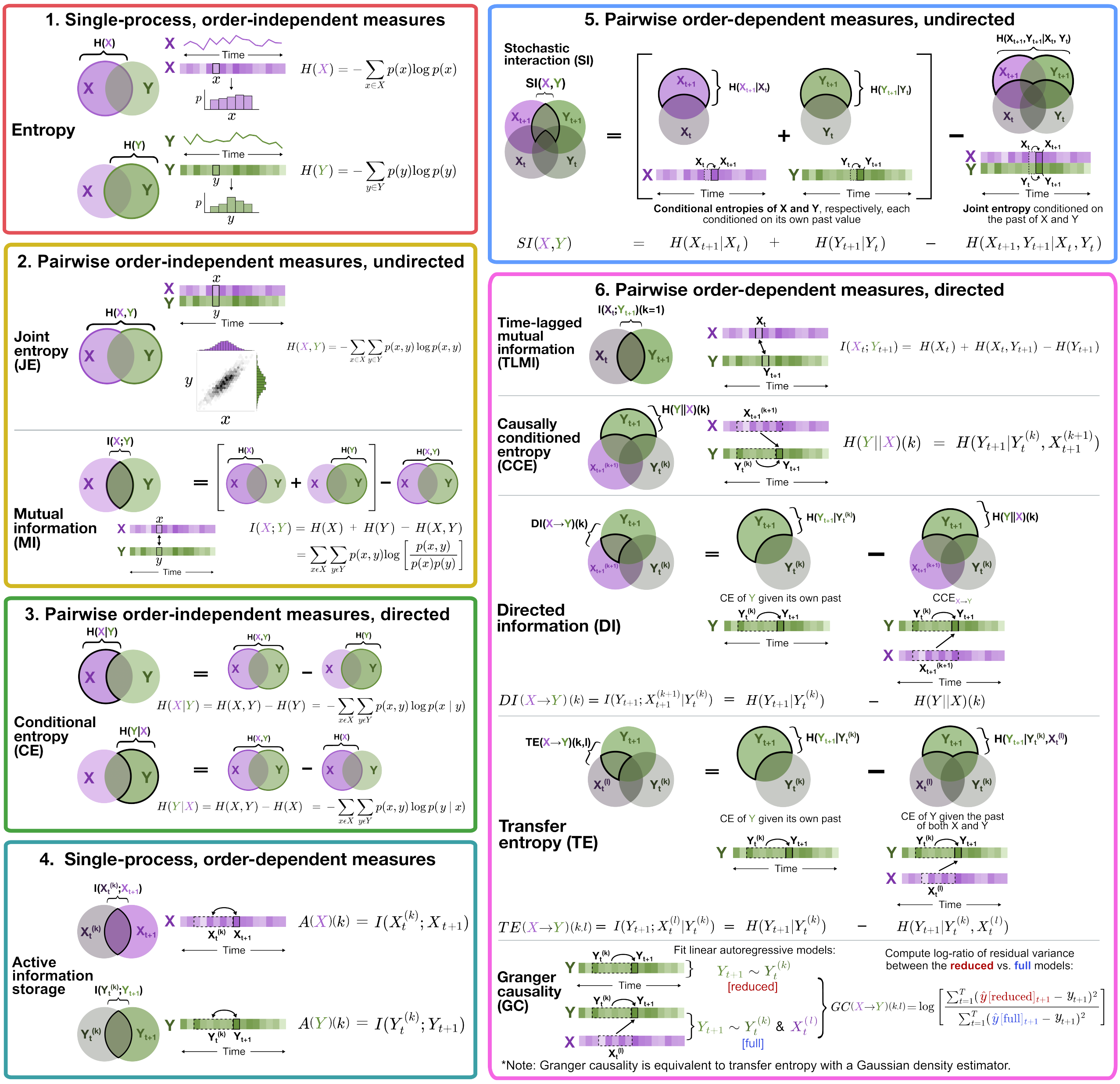

**Figure 3. A unified visual representation of eleven information-theoretic time-series measures, grouped into six categories.** Here, we depict the six broad categories of information-theoretic time series measures (single-process and pairwise) in colored sub-panels, labeled from 1 through 6: (1) single-process, order-independent; (2) pairwise order-independent, undirected (noting that joint entropy of two univariate processes is algorithmically equivalent to a bivariate extension of entropy in the case of a bivariate process); (3) pairwise order-independent, directed; (4) single-process, order-dependent; (5) pairwise order-dependent, undirected; and (6) pairwise order-dependent, directed. Within each sub-panel (e.g., '1. Single-process, order-dependent measures'), all of the individual reviewed measure(s) in that category are visually depicted along with representative mathematical formulae. Venn diagrams are used to illustrate entropy-based concepts for the processes $X$ (shaded in purple) and $Y$ (shaded in green). For the order-dependent measures in sub-panels 4 through 6, we use lighter and less-saturated Venn diagram elements to indicate the past history of a time series, while darker and more saturated elements indicate the present. The eleven reviewed measures are presented in the same order as in Table 1, which contains a high-level summary for each measure along with the section number in which each measure is described in detail.

In this work, we present a unified schematic and tabular overview of eleven core information-theoretic measures used in time-series analysis. Collectively, this synthesis provides direct links between conceptual theory and the implementation and interpretation with empirical data in an applied setting. By juxtaposing widely used measures like mutual information or transfer entropy with lesser-known yet powerful tools, like stochastic interaction or directed information, we introduce a schematic that reveals a compact landscape of how information flows can be quantified across time and between signals. We implement categorization according to the three core characteristics summarized in Fig. 2, reflecting a distilled structure underlying much of the literature—a structure that may otherwise be obscured by disciplinary silos or differences in naming conventions. From these three characteristics, we identified six broad groups of measures:

1. Single-process, order-independent measures;
2. Pairwise order-independent measures, undirected;
3. Pairwise order-independent measures, directed;
4. Single-process, order-dependent measures;
5. Pairwise order-dependent measures, undirected; and
6. Pairwise order-dependent measures, directed.

We organized these six categories and their constituent measures within color-coded sub-panels in Fig. 3. For example, '1. Single-process, order-independent measures' only includes one measure (entropy), while '6. Pairwise order-dependent measures, directed' includes five measures. Our aim with this figure is to consolidate a diffuse and disjoint literature into one cohesive visual representation, directly comparing formulations that condition on past and present values of a single process or pair of processes in slightly different ways. We use the popular Venn Diagram schematic (see e.g., Cover and Thomas [2, p. 48]) to illustrate concepts of marginal entropy; joint entropy and mutual information; and conditional entropy, which collectively form the building blocks of more complex pairwise dependency relationships such as stochastic interaction, transfer entropy, and directed information. We note that Venn diagrams provide useful intuition on how sums and differences of these measures are related, though it may be problematic when used for more than three variables since the analogy to set theory is loose, and where the (presumably positive) areas on the diagrams may be associated with negative quantities ([2, p. 75][3, p. 143][17]). With that said, all of the measures considered here (and their areas on the Venn diagrams) are non-negative, and emphasize their purpose to guide intuition on sums and differences of the quantities.

To complement the visual representation in Fig. 3, we have summarized key aspects of each information-theoretic measure covered in this paper in Table 1 below. This table consolidates information about the directionality, temporal dependence, and scope (single-process vs. pairwise) for each measure. The high-level organization introduced in Fig. 3 and Table 1 forms the basis for the rest of this paper, with a detailed breakdown of each measure in subsequent sections. Each section introduces a summary of what the measure quantifies ('**What**'), how the measure is computed ('**How**'), and why that measure is useful in characterizing dynamical structure in the time series ('**Why**'). Importantly, the figure and table invite the reader to view these measures not as isolated techniques, but as variations on a common theme: the reduction of uncertainty about some variable (e.g., $Y$) by conditioning on others (such as $X$, or the past of $Y$, or both). Information theory provides a powerful language for this task through its flexibility in capturing both linear and nonlinear dependencies independently of the generative form of the underlying data. In practice, there are only so many 'sensible' ways to quantify coupling and/or memory in time series data—and Fig. 3 makes these explicit, highlighting the conceptual logic behind each approach. The goal here is to scaffold the rest of this paper's deeper dives with a birds-eye view that not only clarifies distinctions, but also helps the reader see the broader unity in this diversity of methods. This work is designed to introduce key aspects of information-theoretic tools for time series data, with a focus on the construction of each measure and the theoretical relationships between them. Of note, all information-theoretic measures covered here can be computed from empirical data using a combination of two open-source software packages: the *Java Information Dynamics Toolkit* (JIDT) [18] and the *python toolkit of statistics for pairwise interactions* (pyspi) [19, 20].

| Measure | Notation | Direction | Order sensitivity | Plain-language summary of what this measure captures | Section |
|---|---|---|---|---|---|
| Entropy | $H(X)$ | N/A | Order-independent | How surprising (i.e., uncertain) the samples of process $X$ are on average. | 1.1 |
| Joint entropy | $H(X,Y)$ | Undirected | Order-independent | How surprising (i.e., uncertain) the paired observations of two processes, $X$ and $Y$, are on average. Note that joint entropy of two univariate processes is algorithmically equivalent to a bivariate extension of entropy in the case of a bivariate process. | 2.1 |
| Mutual information | $I(X;Y)$ | Undirected | Order-independent | The amount of information that samples of process $(X)$ provide about those of another $(Y)$, and vice versa. | 2.2 |
| Conditional entropy | $H(Y \mid X)$ | Directed | Order-independent | The uncertainty remaining for the observations in one process $(Y)$ after simultaneously observing another $(X)$. | 3.1 |
| Active information storage | $A(X)$ | N/A | Order-dependent | The amount of information provided by the past values of process $X$ about its present value. | 4.1 |
| Stochastic interaction | $\mathrm{SI}(X,Y)$ | Undirected | Order-dependent | How much more uncertain the future values of $X$ and $Y$ become when they are treated as independent ('disconnected') sources of dynamics, rather than as a single coupled system ('fully interconnected'). | 5.1 |
| Time-lagged mutual information | $I(X_n;Y_{n+1})$ | Directed | Order-dependent | The amount of information shared between the present value of one process $(Y)$ and a past or time-lagged instance of another $(X)$. | 6.1 |
| Causally conditioned entropy | $H(Y \| X)$ | Directed | Order-dependent | The amount of uncertainty remaining in the present value of one process $(Y)$ after observing its own past as well as the past and present of another $(X)$. | 6.2 |
| Directed information | $\mathrm{DI}(X \to Y)$ | Directed | Order-dependent | The amount of information provided by the past and present of a source process $(X)$ about the present of a target process $(Y)$, beyond that which is already explained by the past of $Y$ on its own. | 6.3 |
| Transfer entropy | $\mathrm{TE}(X \to Y)$ | Directed | Order-dependent | The amount of information provided by the past of a source process $(X)$ about the present of a target process $(Y)$, beyond that which is already explained by the past of $Y$ on its own. | 6.4 |
| Granger causality | $\mathrm{GC}(X \to Y)$ | Directed | Order-dependent | The additional predictive power that the past of a source process $(X)$ contributes to the present value of a target process $(Y)$, beyond that of a linear model fit on the past of $Y$ on its own. While GC is not explicitly an information-theoretic measure, it is equivalent to TE computed with a Gaussian density estimator and is therefore included within the scope of this review. | 6.5 |

**Table 1. A high-level overview of the eleven information-theoretic time-series measures covered in this review.** For each of the eleven information-theoretic time-series measures, this table summarizes their notation, directionality (for pairwise measures), time-ordering sensitivity, what the measure captures, and the section in which the measure is covered in detail. Measure names are colored with the same palette as in Fig. 3 to denote in which of the six broad categories each measure sits.

## An illustrative case study focusing on functional neuroimaging time series

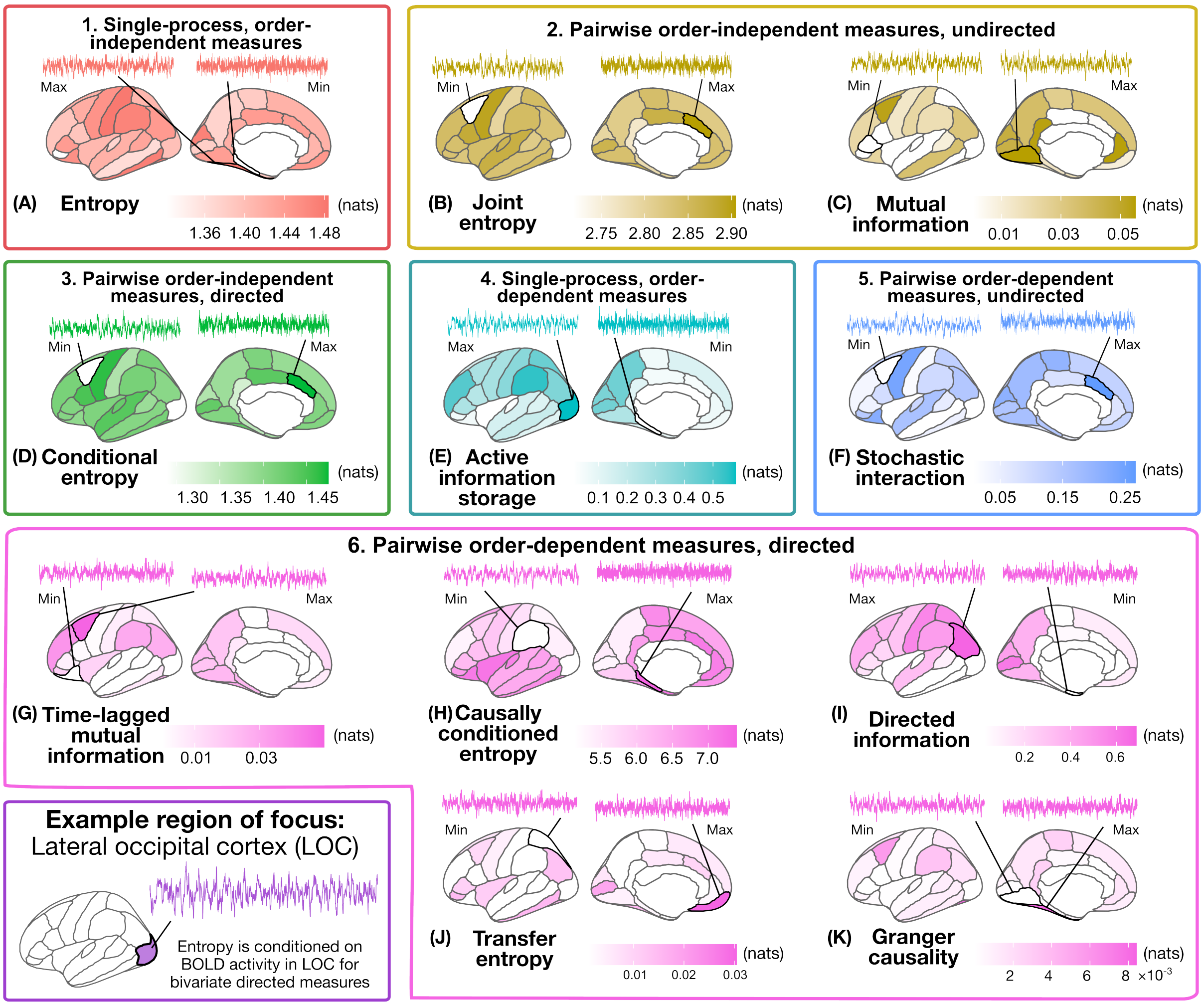

**Figure 4. Visualizing information-theoretic time-series measures on the cortical surface.** To provide a concrete demonstration of the application of information-theoretic time-series statistics, we computed the eleven measures from resting-state BOLD fMRI data from one example participant in the HCP S1200 release [21]. For each of the eleven information-theoretic measures, belonging to six categories in total (see Fig. 3 and Table 1), the resulting measure values are projected onto the left cortical surface of the brain. In the case of the two single-process measures, entropy and active information storage, the values are computed and shown for each individual region. For the nine pairwise measures, we set the lateral occipital cortex (LOC) as the seed region (shown in purple at the bottom left) and computed the pairwise measures to each of the other $33$ regions in the left cortex; the values shown for each brain region reflect the value from the lateral occipital cortex to the corresponding region for the indicated measure. For all measures, we highlight the region with the minimum and maximum value and depict the corresponding fMRI time series.

As an illustrative case study, we focus on functional neuroimaging—comprised of multivariate time series recordings of the activity in various brain areas—to provide context for the implementation and interpretation of each measure covered in the present work. Indeed, there are myriad applications for information theory in neuroscience across both recording scales and modalities, including neuronal spike trains [22, 23], local field potentials [24], and macroscale functional magnetic resonance imaging (fMRI) dynamics [25, 26]. We include a simple example of how and why each measure might be used to map information structure, examining the resting-state blood oxygen level-dependent (BOLD) fMRI time series from one participant in the Human Connectome Project [21] S1200 release (ID ‘298051’, selected at random). For each of the eleven information-theoretic time-series measures depicted in Fig. 3 and summarized in Table 1, the computed values are projected onto the left hemisphere cortical surface in Fig. 4, with one brain map per measure. We maintain the same color-coding and nomenclature as in Fig. 3 and Fig. 4 for consistency, and draw upon the spatial distribution of these eleven distinct measures in the discussion of why each measure is useful for computational neuroscience in particular.

The spatial distributions in Fig. 4 provide intuitive examples of the kinds of signal dynamics each measure captures, contributing to a foundation of deeper theoretical and practical discussion of each measure. This demonstration is purely to showcase the computation and interpretation of the eleven information-theoretic measures covered in this review—rather than drawing any biological conclusions based on a single-participant exemplar. All fMRI data were preprocessed as per earlier work [27] and further described in 'Methods: Examples using empirical neuroimaging data from the Human Connectome Project'.

With the exception of Granger causality (GC), all measures were computed using $k$-nearest-neighbor-based estimators for continuous entropy; these estimators are discussed in the sections 'Entropy' and 'Mutual Information'. While systematic quantitative comparison across estimators (and participants) is beyond the scope of this illustrative case study, we recognize the importance of selecting appropriate measures and estimators for a given dataset. Since many BOLD fMRI studies assume a Gaussian covariance structure [28], as a point of comparison, we also computed these measures across the left-hemisphere cerebral cortex using the Gaussian density estimator (cf. Fig. S1). Qualitatively, the resulting spatial patterns were broadly similar to those obtained with the model-general estimators; however, the Gaussian estimator yielded a systematically compressed dynamic range, consistent with the stronger parametric assumptions imposed on the underlying distribution. In the case of entropy, the Gaussian estimator produced the single constant value of 1.419 nats (these units are discussed more in the 'Entropy' section below), which is precisely the entropy of a Gaussian distribution with variance $\sigma = 1$, corresponding to the $z$-score normalization applied to regional BOLD fMRI time series. Altogether, to provide context for users interested in applying these information-theoretic methods to empirical datasets, we summarize the computation times for each measure (and given estimator) in Fig. S2 and Table S1.

Having introduced our classification of the eleven information-theoretic time-series measures through a schematic depiction (Fig. 3) and tabular overview (Table 1), we now turn to cover each measure in more detail, including a focus on our applied fMRI case study. Subsequent sections are intended to serve as in-depth reference material for the reader interested in a particular measure(s). We begin by examining the first class of measures, which are single-process and order-independent—meaning they capture properties of a signal's distribution that do not depend on temporal structure.

# 1 Single-process, order-independent measures

In this section, we introduce measures that capture properties within a single time series and are independent of the temporal ordering. In other words, the values of the time series could be shuffled along the time axis, and the values of these measures would not change, as they capture properties of the underlying distribution.

## 1.1 Entropy

### 1.1.1 What (does this measure capture):

Entropy—by which we refer to 'Shannon entropy' [1]—measures the degree to which the observed values from our time series, from which the probability distribution is composed, are surprising. More formally, the entropy of the distribution of a given process is an order-independent measure of the apparent randomness or variability exhibited across time series values, without considering temporal structure or other variables [3]. From another perspective, it can be interpreted as the average uncertainty in predicting the value of this variable, again without taking other information into account.

### 1.1.2 How (is this measure computed):

Let $X$ be a random variable representing the values taken at individual time points in a time series (i.e., the distribution over time). As schematically depicted in Fig. 5, the entropy—with notation $H(X)$—is then defined as:

$$H(X) = -\sum_{x \in X} p(x) \log p(x) \,, \tag{1}$$

which takes a weighted sum (over the probability $p(x)$) of the surprise or Shannon information content, $-\log p(x)$ [1], for each possible value $x$ that the random variable $X$ may take over time. Here, the $\log$ may be taken with any base, with the base simply determining the units of the measure. For discrete-valued variables, base 2 is typically used, in which case the units are in 'bits' [1]. Alternatively—by default when no base is indicated, and more often for continuous-valued variables (as defined in the next section)—one may use the natural logarithm with base $e$ (such that the units are therefore in 'nats' [1]). For the remainder of this piece, we will work with the natural logarithm in units of 'nats'. The Shannon entropy, as defined in Eq. (1), is non-negative. When applied to a single variable out of a multivariate set, the measure may be referred to as 'marginal entropy', indicating the marginalization of the joint PDF applied to obtain each $p(x)$ value.

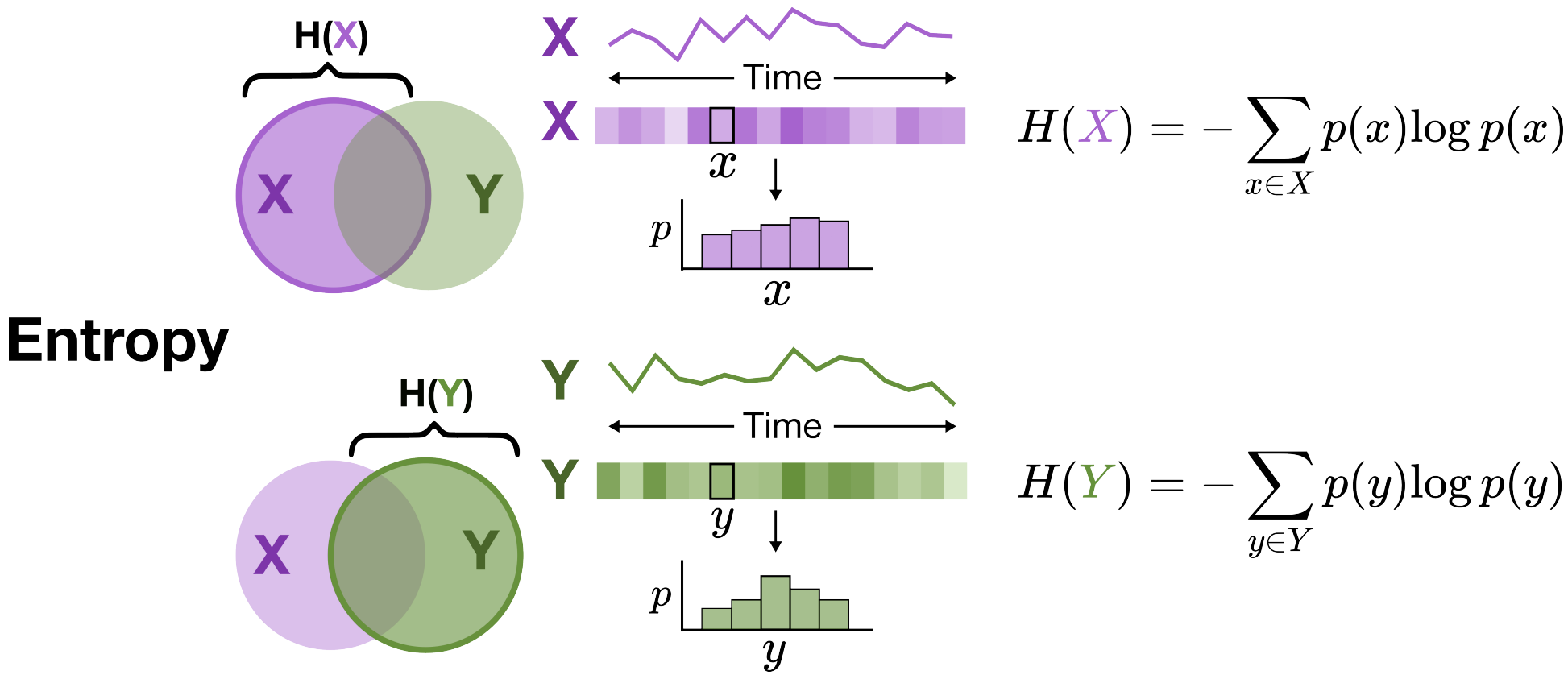

**Figure 5. Entropy.** Entropy quantifies uncertainty in probability distribution for the univariate process $X$ or $Y$, with higher values indicating greater unpredictability in the measured time series. The histograms for the probability distributions across realizations ($x$ and $y$) demonstrate the order-independent nature of entropy computation. The mathematical form for $H(X)$ and $H(Y)$ is provided on the right and discussed in Sec. 1.1.2.

#### An important note on computing entropy in continuous space

Here, we briefly introduce the concept of differential entropy for continuous-valued processes, which forms the basis for our simple illustrative fMRI analysis. When there are continuous variables in a given process (or pair of processes), these measures can be derived from 'differential entropy'—a measure generalizing the Shannon entropy to continuous variables—given as [2, Chapter 8]:

$$H(X) = -\int_{-\infty}^{\infty} f(x) \log f(x) dx \,, \tag{2}$$

for a given random variable $X$, from which individual realizations are denoted with the lower-case $x$. Within the integral, the probability density function[4] of $x$ ($f(x)$) weights the log-probability ($\log f(x)$), such that the differential entropy is an expectation function (like Shannon entropy). The limits of this integral range from $[-\infty, \infty]$ in general, since each realization corresponds to a continuous random variable. The differential entropy has different properties from the Shannon entropy [2, Chapter 8]; most notably, it can be a negative quantity, which means the continuous distribution is more concentrated than the uniform random distribution on $[0, 1]$.

[4] Notice that here we use $f(x)$ to capture the probability density when $X$ is a *continuous* random variable, while $p(x)$ is the notation for the probability of a specific value from a *discrete* random variable.

When we compute information-theoretic measures from empirical data, we obtain an estimate ($\hat{H}$) of an underlying true value ($H$). This requires us to, implicitly or explicitly, estimate the PDF of the distribution from which the time series data in realizations $x$ of $X$ are drawn. The *JIDT* (version 1.6.1) library for computing information-theoretic measures estimates the PDF and approximates the integral in Eq. (2) above, using one of several possible methods listed below which are applicable to continuous-valued variables:

- Gaussian density estimator (`gaussian`)
- Box kernel density estimator (`kernel`)
- Kozachenko–Leonenko density estimator (`kozachenko`) [29]

Each estimator comes with its own assumptions, benefits, and drawbacks. The simplest Gaussian density estimator models the PDF as, unsurprisingly, a Gaussian distribution with variance $\sigma^2$, and under this model, the entropy is given by:

$$H(X) = \frac{1}{2}\log 2\pi e\sigma^2, \tag{3}$$

obtained by plugging in the PDF of the Gaussian distribution $X \sim \mathcal{N}(\mu, \sigma^2)$ for $f(x)$ in Eq. (2). The more complex box kernel and Kozachenko–Leonenko estimators are, in contrast, 'model-general', and approximate probability densities $f(x)$ by counting other samples within various radii of each given sample $x$, and use these (plus perhaps statistical corrections) to then estimate entropy. We refer the interested reader to Lizier [18] and Cliff et al. [20] for a detailed discussion of the implementation and interpretation of each method.

### 1.1.3 Why (is this measure useful):

Having defined the core algorithmic underpinnings of entropy, we continue to explore why it is a useful measure, with applications focused on computational neuroscience. If $X$ represents the activity of a given brain region, a larger $H(X)$ indicates that the activity magnitude values (e.g., BOLD fMRI signal, calcium imaging, etc.) are closer to a more 'spread out' and equiprobable distribution, so each value is more surprising on average. By contrast, a smaller $H(X)$ suggests less diversity in the distribution of values in the given realization $x$ of $X$ (i.e., the observed period of that brain region) in favor of a few dominant values.

Entropy has gained a foothold in computational neuroscience to index the predictability of localized dynamics across spatial and temporal imaging scales [30–32]. We turn to our simple case study example in Fig. 4A, in which we computed the entropy from resting-state BOLD fMRI time series for all brain regions in the left cerebral cortex of one individual (see 'Methods: Examples using empirical neuroimaging data from the Human Connectome Project' for details). As shown on the cortical surface, we observe a narrow range of entropy values across different cortical regions computed with the Kozachenko–Leonenko estimator. This range spans from 1.32 nats in the parahippocampal gyrus to 1.48 nats in the fusiform gyrus, indicating generally homogeneous entropy values across the left cortex in this individual—such that all measured resting-state BOLD time series are comparably concentrated around the mean and contain comparable amounts of surprisal in this individual. The entropy of a single-process time series, which is insensitive to the time ordering, serves as a building block for more complex algorithmic quantities (namely, mutual information and conditional entropy), which are introduced in subsequent sections.

# 2 Pairwise order-independent measures, undirected

## 2.1 Joint entropy

### 2.1.1 What:

Joint entropy (JE) [3] quantifies the average information content—or 'surprisal'—across all paired observations of two random variables, $X$ and $Y$, over their full joint distribution. It reflects the total uncertainty associated with jointly observing both $X$ and $Y$, without assuming any particular relationship between them.

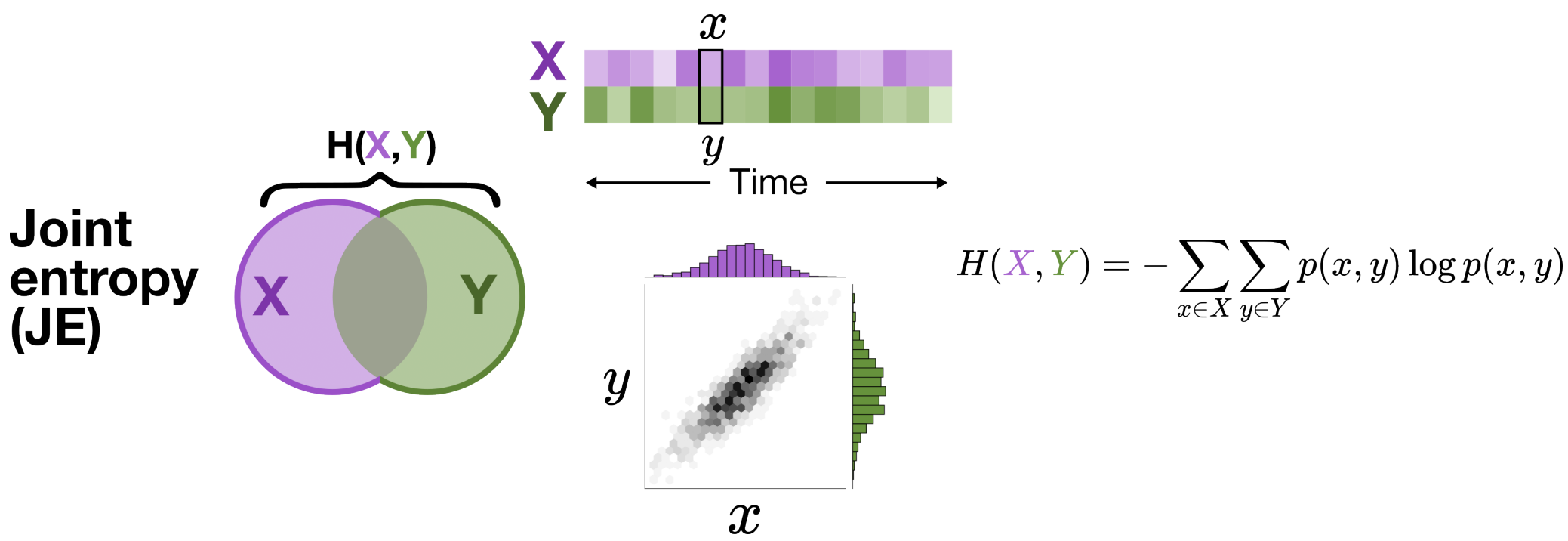

**Figure 6. Joint entropy.** Joint entropy (JE) quantifies uncertainty in the joint probability distribution for processes $X$ and $Y$. The hexagonal heatmap of two-dimensional bin counts depicts the density of values in $x$–$y$ space, demonstrating the invariance to temporal shuffling as a measure that does not depend on temporal order. The mathematical form for $H(X,Y)$ is provided on the right and discussed in Sec. 2.1.2.

### 2.1.2 How:

As schematically depicted in Fig. 6, the joint entropy between random variables $X$ and $Y$ is defined as

$$H(X,Y) = -\sum_{x\in X}\sum_{y\in Y} p(x,y)\log p(x,y)\,, \tag{4}$$

where $x \in X$ and $y \in Y$ indicate the samples spaces of $X$ and $Y$, and $p(x,y)$ is the joint probability mass function. This definition generalizes the single-process entropy (cf. Sec. 1.1) by considering the joint distribution of *two* variables rather than a single one. Strictly speaking, joint entropy would belong to our 'single-process' category; however, here we classify it as 'pairwise' in order to focus on how it can be used to specifically capture the relationship *between a pair* of variables. As with single-process entropy, joint entropy is *not* an order-dependent measure. Of course, this can be fully generalized to any number of joint variables for $H(\boldsymbol{X})$; where the boldface $\boldsymbol{X}$ denotes a vector comprising a multivariate time series with $\{X_1,\ldots,X_M\}$ as individual variables corresponding to individual processes, and $H(\boldsymbol{X})$ is defined as a corresponding function of the multidimensional probability distribution function $p(\boldsymbol{x}) = p(x_1,\ldots,x_M)$.

Similarly, we can extend the single-process differential entropy introduced in Sec. 1.1 to the multivariate case here:

$$H(\boldsymbol{X}) = H(X_1,\ldots,X_M) = -\int_{-\infty}^{\infty} f(x_1,\ldots,x_M)\log f(x_1,\ldots,x_M)dx\,, \tag{5}$$

where $f(x_1,\ldots,x_M)$ is the joint probability density function (PDF) of the continuous random vector $\boldsymbol{X} = (X_1,\ldots,X_M)$. The same estimators described in Sec. 1.1 are available for joint entropy—namely, Gaussian, box kernel, and Kozachenko–Leonenko. There is a special case for the Gaussian density estimator: for the multivariate normal distribution $X \sim \mathcal{N}(\mu,\Sigma)$ of dimension $M$, the differential entropy has the following closed-form expression (i.e., no integral needs to be numerically evaluated):

$$H(\mathbf{X}) = \frac{1}{2}\log{(2\pi e)^M}|\Omega|, \tag{6}$$

given the covariance matrix $\Omega$ (with determinant $|\Omega|$) between its $M$ sub-variables $f(x_1,\ldots,x_M)$. The intuition behind this for Gaussian multivariate distributions is that the larger the covariance among $M$ dimensions, the more 'spread out' the distribution—and therefore, the entropy is greater. If all the $M$ variables in $\boldsymbol{X}$ are independent, $\Omega$ is diagonal (i.e., all pairwise covariances are zero), and the joint entropy becomes the sum of the individual entropies.

In empirical settings, such as when working with two observed time series, this theoretical quantity must be estimated from finite samples. $H(X,Y)$ operates in the two dimensions of realizations $x$ and $y$ (with the underlying probability distribution represented as the heatmap, using hexagons to indicate two-dimensional bin counts, in Fig. 6), summing over the product of the joint probability and the log-transformed joint probability for each pair of observations in $x$ and $y$.

### 2.1.3 Why:

If $X$ and $Y$ represent the activity of two brain regions, the joint entropy $H(X,Y)$ is a measure of the shared variability and combined uncertainty between the two regions—providing insight into the nature of their functional coupling. A larger $H(X,Y)$ indicates that the two brain regions collectively exhibit greater diversity or variability in their combined states (i.e., across $x-y$ space in the heatmap in Fig. 6). This means that, on average, each pair of $x,y$ activity values is more surprising than if the time series were dominated by a small handful of repeated values, which would result in a more concentrated distribution in $x-y$ space. The inequality $H(X,Y) \leq H(X)+H(Y)$ formalizes that the JE of $X$ and $Y$ is maximized only when $X$ and $Y$ are statistically independent—such that any deviation from this upper bound of $H(X)+H(Y)$ reflects shared structure, or redundancy, between $X$ and $Y$.

In our simple example in Fig. 4B, we compare the pairwise joint entropy (JE) between an example brain region of focus (lateral occipital cortex, LOC) and all other regions in the left cortex, considered one-by-one, using the Kozachenko–Leonenko estimator. Here, we observe the greatest JE between the caudal anterior cingulate cortex and the LOC, which might suggest that the caudal anterior cingulate adds maximum diversity to the LOC in this example fMRI dataset out of all regions evaluated. This implies that the joint behavior of these two regions is likely the least constrained by a fixed relationship out of all evaluated regions, indicating more independent or less coupled activity between the two regions. By contrast, a smaller $H(X,Y)$ in comparison to $H(X)+H(Y)$ means a stronger indicated dependency between measured activity in regions $X$ and $Y$. In this case, the caudal middle frontal cortex shows the lowest JE with the LOC, which could reflect redundancy in the information they encode.

While JE has not been directly applied as extensively in computational neuroscience compared to other measures covered here, JE has yielded insights into inter-areal coupling from both fMRI and electroencephalography (EEG) time series. For example, Baseer et al. [33] computed JE to predict microsleep periods from EEG recordings, positing that its strong performance reflects its ability to thoroughly capture how two regions jointly traverse across states. Martin et al. [34] used JE as part of a maximum entropy estimation framework, analyzing fMRI data from Watanabe et al. [35] to show that functional brain architecture is predominantly shaped by pairwise inter-areal coupling (rather than higher-order interactions). As with single-process entropy, JE is a component of more algorithmically complex measures—namely, mutual information (Sec. 2.2), conditional entropy (Sec. 3.1), and stochastic interaction (Sec. 5.1).

## 2.2 Mutual information

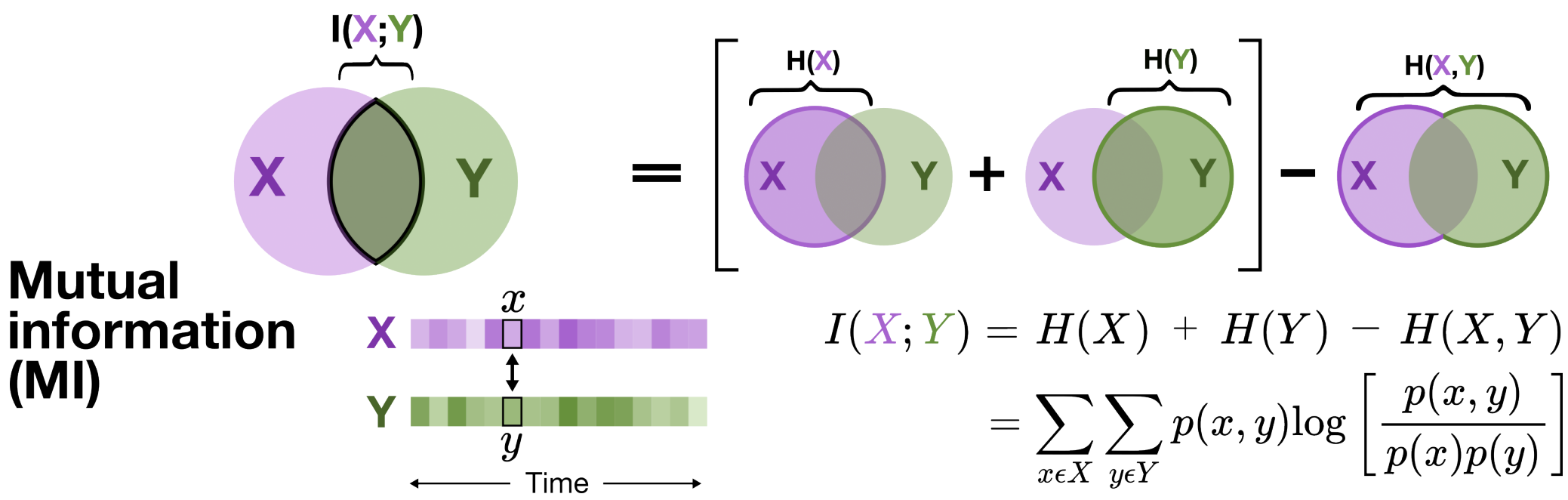

**Figure 7. Mutual information.** Mutual information (MI) quantifies contemporaneous information shared between processes $X$ and $Y$—specifically for the contemporaneous bivariate case here. The Venn diagrams illustrate that this quantity is equal to the sum of the individual entropies, $H(X)$ and $H(Y)$, minus the joint entropy, $H(X,Y)$, which is also reflected in the mathematical form below the diagrams and discussed in Sec. 2.2.2.

### 2.2.1 What:

Mutual information (MI) [3] is an undirected measure of the average information that the value of one time series (e.g., $X$) provides about another (e.g., $Y$), denoted $I(X;Y)$. This can also be phrased in that $I(X;Y)$ tells us how much knowing $X$ reduces uncertainty about $Y$ on average, and vice versa.

### 2.2.2 How:

MI is schematically depicted in Fig. 7 and given by

$$I(X;Y) = H(X) + H(Y) - H(X,Y), \tag{7}$$

$$= \sum_{x \epsilon X} \sum_{y \epsilon Y} p(x,y) \log \left[ \frac{p(x,y)}{p(x)p(y)} \right], \tag{8}$$

which defines $I(X,Y)$ as the difference between the sum of the individual entropies, $H(X)$ and $H(Y)$, minus the joint entropy, $H(X,Y)$ [3] in Eq. (7). Equivalently, Eq. (8) formulates MI directly in terms of probabilities of $X$ and $Y$, summed across all possible values $x \in X$ and $y \in Y$, where $p(x,y)$ is the probability of the paired observations $x$ and $y$ and $p(x)$ and $p(y)$ are the marginal probabilities for $x$ and $y$, respectively. $I(X;Y) \geq 0$, with the $I(X;Y) = 0$ case reflecting total independence between $X$ and $Y$—such that knowing $X$ does not reduce the uncertainty about $Y$ at all, and vice versa. The formulation of MI in Eq. (8) means it is a type of measure known as a Kullback–Leibler (KL) divergence [2, 36]. Consequently, Eq. (8) measures the deviation of the joint distribution $p(x,y)$ from the reference case $p(x)p(y)$, representing what the joint distribution would be if $X$ and $Y$ were independent.

While largely beyond the scope of this review (which focuses on single-process and pairwise measures), we note that MI can be extended to 'conditional mutual information', which conditions the MI between $X$ and $Y$ on knowledge of a third variable, $Z$. Put simply, this involves making the relevant probability functions in Eq. (8) conditional on $z$ and then averaging over $p(z)$, or formulating the conditional MI, for example, as:

$$I(X;Y|Z) = I(X;\{Y,Z\}) - I(X;Z), \tag{9}$$

where $I(X;\{Y,Z\})$ is the MI between variable $X$ and the joint variable $\{Y,Z\}$. We refer the interested reader to Cover and Thomas [2, Chapter 2] for further details and interpretations.

**A note on estimating MI for continuous-valued variables**

For continuous variables, the MI can be estimated as the sums and differences of the individual entropy terms in Eq. (7). Importantly, unlike differential entropy, MI for continuous variables is non-negative (much like the MI for discrete variables), such that MI has a consistent interpretation regardless of the variable type. While MI is generally a nonlinear measure, when we approximate the PDF with a Gaussian distribution (as with the Gaussian density estimator in *JIDT* [18]), it can be derived directly from the Pearson product-moment correlation between $X$ and $Y$, $r$, as

$$\mathrm{MI}_{\mathrm{Gaussian}} = -\frac{1}{2} \log(1 - r^2). \tag{10}$$

This MI implementation assumes that the joint distribution $p(x,y)$ is bivariate Gaussian, with linear interactions between $X$ and $Y$. As such, $\mathrm{MI}_{\mathrm{Gaussian}}$ is only sensitive to linear dependence between $X$ and $Y$. However, the other estimators introduced thus far (i.e., box kernel and Kozachenko–Leonenko) do not make parametric assumptions about the underlying joint bivariate distributions, and can therefore capture both linear and nonlinear relationships between $X$ and $Y$. There is an additional estimator available based on the method in Kraskov et al. [37]—referred to as Kraskov–Stögbauer–Grassberger or KSG (and denoted as `kraskov` in *pyspi* and *JIDT*)—building on the Kozachenko–Leonenko estimator with specific statistical corrections tailored for MI [38]. Specifically, the Kraskov–Stögbauer–Grassberger estimator uses nearest-neighbor statistics (rather than kernel density estimation) to estimate entropy, thereby combining multiple Kozachenko–Leonenko estimators to mitigate bias. The estimator(s) available for each of the eleven information-theoretic estimators discussed here are summarized below in Table 2. We refer the interested reader to Wibral et al. [39] and Vicente and Wibral [40] for more discussion of the Kraskov–Stögbauer–Grassberger method as a suitable 'hybrid' estimator for MI.

**Table 2. Overview of continuous entropy estimators available for each of the eleven information-theoretic measures.** For each measure, the available estimator(s) as implemented in *JIDT* [18] and/or *pyspi* [20] are indicated with a 'Y'. The 'Y*' for transfer entropy (TE) and Granger causality (GC) reflects the fact that these measures are mathematically equivalent using a Gaussian density estimator.

| **Measure** | **Gaussian** | **Kozachenko-Leonenko** | **KSG** |
|---|---|---|---|
| Entropy (ENT) | Y | Y | |
| Joint entropy (JE) | Y | Y | |
| Mutual information (MI) | Y | | Y |
| Conditional entropy (CE) | Y | Y | |
| Active information storage (AIS) | Y | | Y |
| Stochastic interaction (SI) | Y | Y | |
| Time-lagged mutual information (TLMI) | Y | | Y |
| Causal conditioned entropy (CCE) | Y | Y | |
| Directed information (DI) | Y | Y | |
| Transfer entropy (TE) | Y* | | Y |
| Granger causality (GC) | Y* | | |

### 2.2.3 Why:

If $X$ and $Y$ represent the activity of two brain regions, $I(X;Y)$ captures the extent to which knowing the activity of one brain region ($X$) reduces the uncertainty (i.e., improves the predictability) of the other region ($Y$), and vice versa. A larger $I(X;Y)$ indicates that the two regions share more information, implying a stronger inter-region dependence. By contrast, a smaller $I(X;Y)$ points to more independent observations in $X$ and $Y$, with the two regions sharing minimal information with each other. The similarity between MI and the Pearson correlation coefficient (becoming a direct relationship with a Gaussian estimator in Eq. (10)) directly suggests MI as an alternate measure for functional connectivity [41]. In contrast to the Pearson correlation, though, the dependency captured by MI can extend beyond linear relationships, including forms like a function mapping activity of region $X$ to region $Y$ or complex patterns, such that $I(X;Y)$ can capture both linear and nonlinear dependence. This ability to capture both linear and nonlinear relationships, together with the inherently model-general nature, is a strength of MI for measuring inter-areal functional connectivity.

In our simple example in Fig. 4C, we compare the MI between the lateral occipital cortex (LOC) as our region of focus and all other regions in the left cortex using the Kraskov–Stögbauer–Grassberger estimator. Here, we see the greatest MI between the left lingual gyrus and LOC, indicating that knowing the BOLD fMRI time series in the left lingual gyrus maximally reduces uncertainty in that of the LOC out of all regions we evaluated (and vice versa). By contrast, the pars triangularis exhibits the lowest MI with the LOC, which indicates that the LOC shares the least information with the pars triangularis relative to all other regions in this example fMRI dataset. Quantifying inter-regional MI has been used to link microscopic and macroscopic network properties in the macaque functional connectome [42] and to distinguish pediatric epilepsy cases from age-matched controls using BOLD fMRI [43]. MI has also proven useful in characterizing region-specific brain dynamics in response to different sensory stimuli across modalities, including EEG, fMRI, and magnetoencephalography (MEG) [15, 44].

# 3 Pairwise order-independent measures, directed

## 3.1 Conditional entropy

### 3.1.1 What:

Conditional entropy (CE) [3] quantifies the uncertainty in the observations of one process, $Y$, in the context of simultaneously observing another process, $X$. CE is a directed measure, with the CE of $Y$ given $X$—denoted as $H(Y \mid X)$—capturing the amount of remaining uncertainty in $Y$ after conditioning on the observations of $X$. In other words, $H(Y \mid X)$ tells us how much information is still needed to describe $Y$ even when we know the observed values of the conditional variable $X$. As with entropy, joint entropy, and mutual information, the conditional entropy is *not* dependent on the order of values in $X$ and $Y$, making it an order-independent measure.

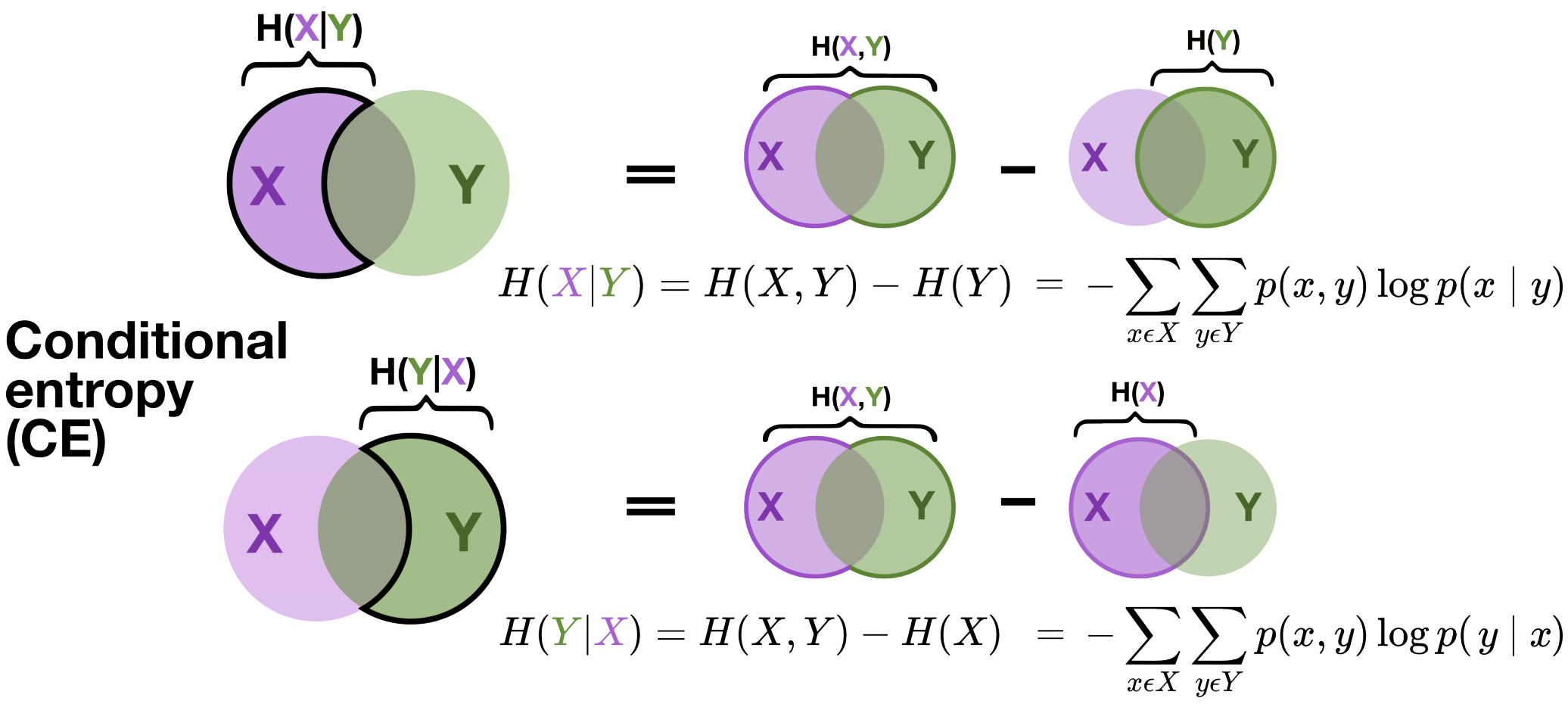

**Figure 8. Conditional entropy.** Conditional entropy (CE) quantifies uncertainty remaining in the probability distribution of one process ($X$) after conditioning on that of another process ($Y$). Both $H(X \mid Y)$ and $H(Y \mid X)$ are shown as this is a directed, and therefore asymmetric, measure. In both cases, the CE (e.g., $H(X \mid Y)$) is equivalent to the joint entropy ($H(X,Y)$) minus the marginal entropy of the process upon which the CE is conditioned ($H(Y)$). The mathematical form is provided for both $H(X \mid Y)$ and $H(Y \mid X)$ and discussed in Sec. 3.1.2.

### 3.1.2 How:

As depicted in Fig. 8, $H(Y \mid X)$ is given as

$$H(Y \mid X) = H(X,Y) - H(X), \tag{11}$$

$$= H(Y) - I(X;Y), \tag{12}$$

$$= \sum_{x \epsilon X} \sum_{y \epsilon Y} p(x,y) \log \left[ \frac{p(x)}{p(x,y)} \right], \tag{13}$$

$$= -\sum_{x \epsilon X} \sum_{y \epsilon Y} p(x,y) \log p(y \mid x), \tag{14}$$

with Eq. (11) showing that $H(Y \mid X)$ is computed as the joint entropy, $H(X,Y)$, minus the marginal entropy of $X$, $H(X)$. This is mathematically equivalent to computing the difference between the marginal entropy of $Y$, $H(Y)$, minus the MI $I(X;Y)$, as in Eq. (12). In practice, for continuous-valued variables, CE is estimated as the difference between two differential entropies for the joint distribution $p(x,y)$, with the differential entropy equation given in Eq. (5), minus the univariate distribution $p(x)$, as given in Eq. (2).

Intuitively, the resulting quantity given by $H(Y|X)$ is the area of the Venn Diagram occupied exclusively by $Y$ in Fig. 8, corresponding to the remaining uncertainty in $Y$ after conditioning on $X$. It follows that $H(Y \mid X)$ can therefore be equivalently written, as in Eq. (14), as the average uncertainty remaining in the observed realization $y$ after conditioning on the observed realization $x$, for all $x \in X$ and all $y \in Y$. Here, as in MI (cf. Sec. 2.2), $p(x,y)$ is the probability of the paired observations $x$ and $y$, and $p(x)$ and $p(y)$ are the marginal distribution probabilities for $x$ and $y$, respectively. The form of the conditional entropy as the difference between joint minus marginal entropy in Eq. (11) can be seen as a direct consequence of the expansion of $p(y|x)$ (the probability distribution over which we are computing entropy here) from Eq. (14) using Bayes' rule. Since CE is a directed measure, the amount of uncertainty remaining for $Y$ after simultaneously observing $X$ is not necessarily equivalent to that remaining in $X$ after observing $Y$. While the Venn diagram circles in Fig. 8 are symmetric (for aesthetic purposes), the two processes can exhibit marginal entropies ($H(X)$ and $H(Y)$) of different magnitudes in general—yielding two different values when subtracting the marginal entropy from the joint entropy.

#### 3.1.3 Why:

If $X$ and $Y$ represent the activity of two brain regions, $H(Y \mid X)$ measures how much uncertainty remains for the distribution of activity values in region $Y$ given knowledge of the activity in region $X$. Practically speaking, conditional entropy treats each time point in the two brain regions individually without actually capturing any temporal dependence, such that it does not capture any element of forward prediction in time. A larger $H(Y \mid X)$ indicates that simultaneously observing the activity in region $X$ does not reduce the uncertainty in region $Y$ by much. This exhibits a complementary relationship to mutual information, for fixed $H(Y)$, as shown in Eq. (12). By contrast, a smaller $H(Y \mid X)$ means that there is little uncertainty remaining about the observations in region $Y$ after simultaneously observing and having access to the activity values of region $X$. In the literature, CE has been applied to EEG [45, 46], fMRI [47, 48], and neural spike train time series [49] to evaluate how knowing information about activity in one brain area (or neuronal ensemble) increases or decreases uncertainty about activity in another brain area.

In our simple example in Fig. 4D, we compute the CE for every cortical region in the left hemisphere (our $Y$ in this example) conditioned on the lateral occipital cortex (LOC, our $X$ in this example), using the Kozachenko–Leonenko estimator—in other words, taking the difference between two differential entropies. For the example of the fMRI activity in the entorhinal cortex, this CE would have the notation H(entorhinal|LOC). Here, we observe the highest CE in the left caudal anterior cingulate cortex, which indicates maximal remaining uncertainty in the caudal anterior cingulate even after observing the LOC activity compared to all other target regions. As noted above, this relationship does not automatically hold in the other direction, from caudal anterior cingulate to LOC; indeed, it is still possible that observing activity in the caudal anterior cingulate could reduce the majority of uncertainty remaining for that of the LOC. This asymmetry is important in understanding the relationship between CE and MI: they would be complementary as above for fixed $H(Y)$ if we used the LOC as $Y$, but not where $Y$ is being changed. This is the reason why CE and MI are not perfectly complementary in this experiment—and therefore why the region with the highest CE after conditioning on the LOC (i.e., the caudal anterior cingulate) is not automatically the region with the lowest MI with the LOC (pars triangularis, cf. Fig. 4C). By contrast, the caudal middle frontal cortex exhibited the lowest CE after conditioning on the LOC, which suggests, for this example fMRI dataset, that observing the BOLD activity in the LOC maximally reduces the uncertainty remaining in the caudal middle frontal region. We do note that these two regions with the minimum and maximum CE are also the same two for JE (cf. Fig. 4B), which only holds as we are computing (entorhinal|LOC) and (caudal anterior cingulate|LOC) rather than the other way around (i.e., activity in the LOC conditioned on the entorhinal cortex or caudal anterior cingulate). In other words, the direct mapping of the maximum and minimum CE and JE in Eq. (11) is dependent on the conditioning directionality, as observed above for the MI.

## 4 Single-process, order-dependent measures

### 4.1 Active information storage

#### 4.1.1 What:

Active information storage (AIS) [50] is a time-dependent measure that captures the amount of information *actively* in use from the past values of a time series to generate the next state. More formally, AIS quantifies information shared between the current time point, $X_{t+1}$, and previous time points up to memory length $k$ (given as $X_t^{(k)}$) for the time series $X$ with length $T$. AIS, referred to as $A(X)$, is the first-order component of a broader measure known as predictive information [51], which quantifies the amount of information about the whole future of a process that can be found in its past—and vice versa.

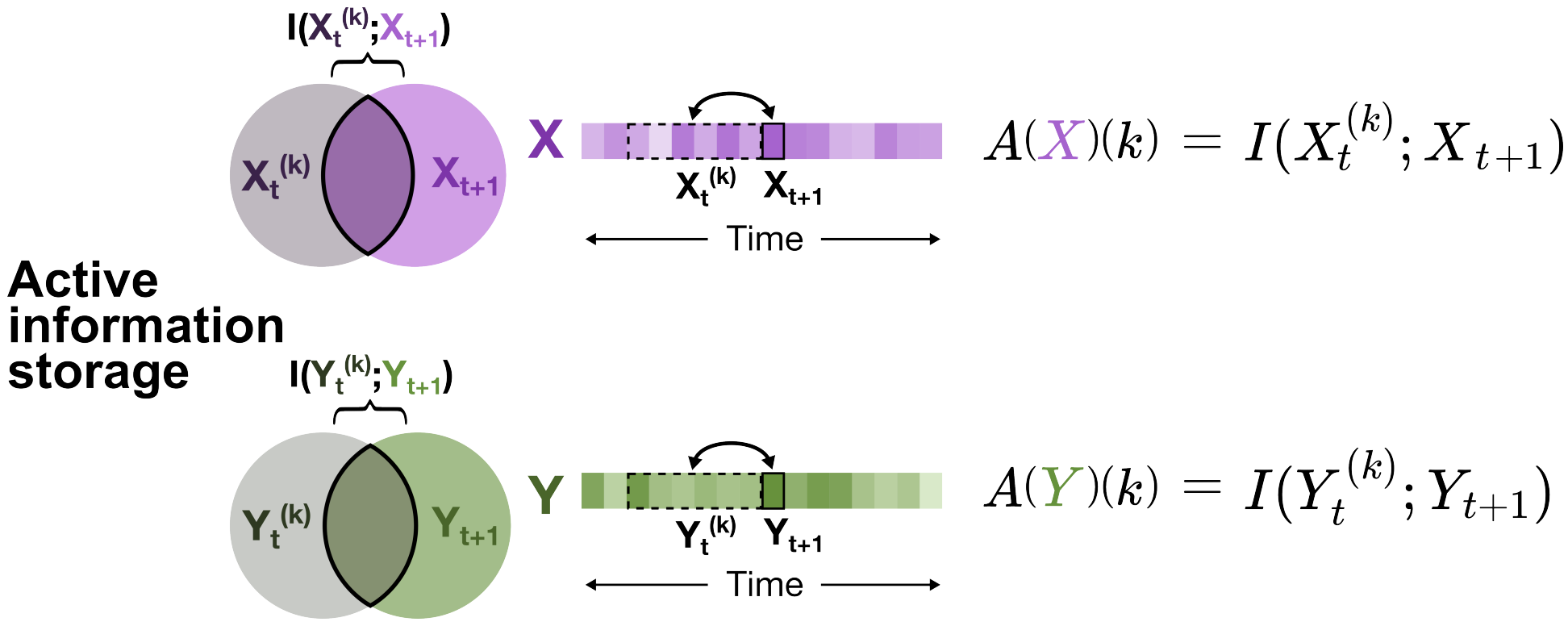

**Figure 9. Active information storage.** Active information storage (AIS) quantifies the amount of information shared, and actively used, between past and future states of a process (e.g., $X$) to predict the future of $X$. This is a single-process, order-dependent measure, equivalent to the mutual information (MI) between the past and present of the process. The mathematical form is provided for both $A(X)(k)$ and $A(Y)(k)$, where $k$ is the memory length, as discussed in Sec. 4.1.2.

### 4.1.2 How:

As schematically depicted in Fig. 9, $A(X)$ is given as

$$A(X)(k) = I(X_t^{(k)}; X_{t+1}), \tag{15}$$

where $t$ is a time point in the range $[1,T]$ and $T$ is the total length of time series $X$. $k$ denotes the memory length (i.e., the number of time points prior to time $t+1$ considered), such that $X_t^{(k)} = \{X_t, X_{t-1}, \ldots, X_{t-k+1}\}$. The expression as $I(X_t^{(k)}; X_{t+1})$ indicates that the AIS is an MI (cf. Sec. 2.2) computed between the present value of $X$ and the past of $X$ up to memory length $k$. The value chosen for $k$ determines the number of past time points examined, with $A(X)(k)$ measuring the information storage under the assumption that the process $X$ is a $k$-order Markov chain.

Theoretically, the window length $k$ can be set to the full length of the time series $X$ ($T$) minus one—meaning that the entire past of $X$ is considered. Empirically, however, evaluating the *entire* past of $X$ can become problematic due to computational burden (which becomes more pronounced in the case of pairwise order-dependent measures, discussed in subsequent sections). Moreover, when we have only a singular time series realization for $X$, we would only have single samples for the vectors $X_t^{(t)}$ at each time $t$. One approach to address these issues is for samples $(x_t^{(k)}, x_{t+1})$ across the whole time series to be utilized via an ergodic assumption to make one AIS calculation, $A(X)(k)$—rather than at each time point, $t$. Optimizing selection of the parameter $k$ usually involves a trade-off between wanting a longer $k$ to capture more memory, while avoiding bias due to a too highly-dimensional probability space (see strategies detailed in [52, 53]). In contrast to a linear autocorrelation profile, $A(X)$ captures both linear and nonlinear dependence, while also capturing the full multivariate dependence of the variable $X$ on its whole past—rather than characterizing the pairwise dependence at each lag separately.

### 4.1.3 Why:

AIS is particularly useful in computational neuroscience to quantify the complexity and predictability of a neural process as it evolves through time, with higher values requiring both dynamical 'richness' (i.e., complex and informative patterns over time) and temporal predictability [54, 55]. If $X$ represents the activity of a given brain region, $A(X)$ captures how much the current activity can be predicted from that region's past activity states. A larger value of $A(X)$ means there is a stronger dependence between the region's past and present activity values—indicating the activity is dynamically rich and highly predictable [55]. By contrast, a smaller $A(X)$ suggests there is weak or no dependence between the region's past and present activity values.

In Fig. 4E, we compare the AIS (using $k=1$) from BOLD fMRI time series in all regions in the left cortex from one example HCP participant using the Kraskov–Stögbauer–Grassberger estimator. Here, we observe the highest AIS in the LOC, suggesting that this region exhibits richer dynamics with more information from past activity values used in future states. By contrast, the parahippocampal gyrus exhibits the lowest overall AIS, meaning its activity is less dependent on its own past states—possibly indicating a greater relative influence of other regions' prior activity on future activity in the parahippocampal gyrus. AIS has been applied to characterize neural dynamics around a critical point [56, 57], learning and memory in neuromorphic computing [58], frequency-specific rhythms during anesthesia [59], and distributed processing of visual stimuli [60, 61]. Prior clinically grounded work has shown that AIS is reduced in the hippocampus of individuals with autism spectrum disorder [55, 62] and in the motor cortex of individuals with Parkinson's disease [63].

# 5 Pairwise order-dependent measures, undirected

## 5.1 Stochastic interaction

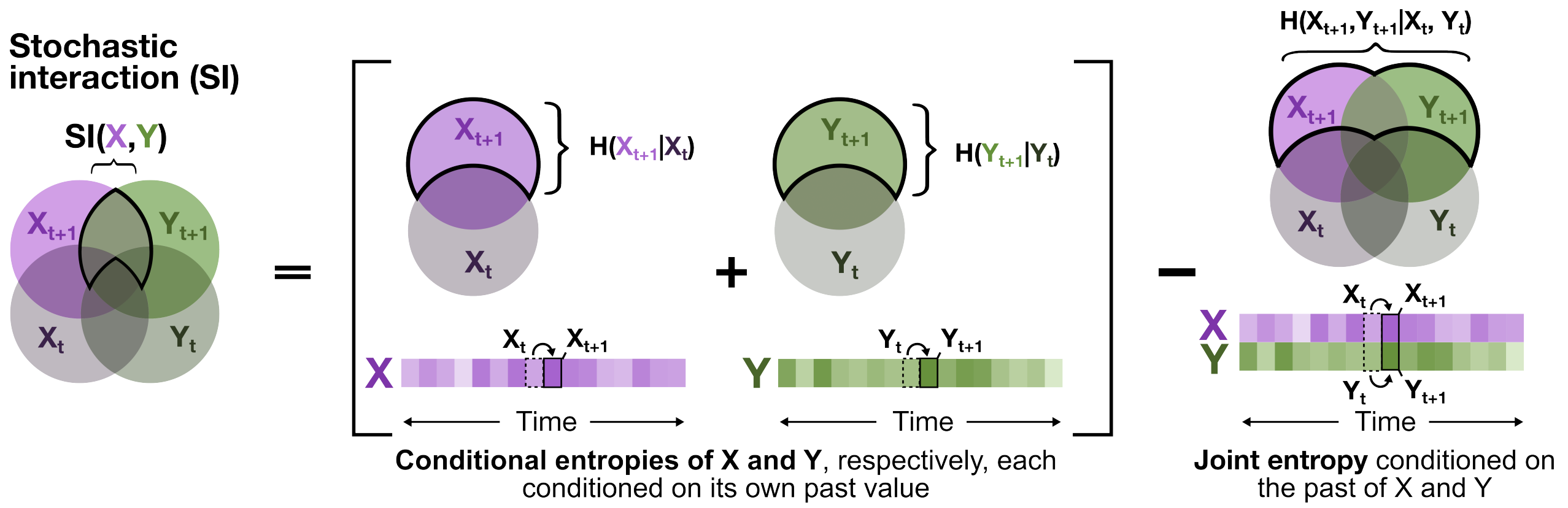

$$SI(X,Y) = H(X_{t+1}|X_t) + H(Y_{t+1}|Y_t) - H(X_{t+1},Y_{t+1}|X_t,Y_t)$$

**Figure 10. Stochastic interaction.** Stochastic interaction (SI) measures the integrated information between the two processes $X$ and $Y$ that cannot be explained by considering the two time series separately. It is an order-dependent pairwise measure that considers the past and present of both $X$ and $Y$. As depicted by the Venn diagrams, $\mathrm{SI}(X,Y)$ is equivalent to the sum of the conditional entropy of $X$ and $Y$ conditioned on their own past values, respectively, minus the joint entropy of $X$ and $Y$ both conditioned on their past values together. The mathematical form is provided below the diagram schematics and discussed in Sec. 5.1.2.

### 5.1.1 What:

Stochastic interaction (SI) [64, 65], denoted $\mathrm{SI}(X,Y)$, quantifies the integrated information between two processes, $X$ and $Y$, by capturing the joint dynamical dependence between the two that cannot be entirely explained by their marginal dynamics. As described in Cohen et al. [66], SI captures the amount of information lost in (modeling) a 'disconnected' system compared to (modeling) an 'integrated' system. Specifically, SI measures the extent to which 'disconnecting' $X$ and $Y$—in other words, modeling them as independent processes—increases the uncertainty about the future values of $X$ and $Y$ based on their respective past values, compared to the uncertainty about future values in the fully 'interconnected' (or jointly modeled) paired system. We previously reported that SI behaves very similarly across data types to measures derived from the integrated information literature [20], such as geometric integrated information and $\phi^*$ [67]. For an in-depth discussion of how SI relates to the broader field of integrated information (particularly geometric integrated information [67]), we refer the interested reader to Mediano et al. [68].

### 5.1.2 How:

As schematically depicted in Fig. 10, $\mathrm{SI}(X,Y)$ is formulated based on Equation #26 in Ay [65]:

$$\mathrm{SI}(X,Y) = H(X_{t+1}|X_t) + H(Y_{t+1}|Y_t) - H(X_{t+1},Y_{t+1}|X_t,Y_t), \qquad \textbf{(16)}$$

where $H(X_{t+1}|X_t)$ indicates the entropy of $X$ at time point $t+1$ conditioned on the previous value at time point $t$—in other words, the amount of uncertainty remaining in the value of $X$ at time $t+1$ after knowing the previous value of $X$ at time $t$—as is the case for $H(Y_{t+1}|Y_t)$. The subtracted term in Eq. (16), $H(X_{t+1}, Y_{t+1}|X_t, Y_t)$, captures the joint entropy of $X$ and $Y$ at time $t+1$, both conditioned on their previous values at time $t$—in other words, the amount of uncertainty remaining in the paired observation $X_{t+1}, Y_{t+1}$ after conditioning on the past paired observation $X_t, Y_t$. We note that the definition of SI here considers only one past value of $X$ and $Y$, akin to using $k = 1$ for the AIS from the previous section. This effectively assumes a first-order joint Markov process, although in principle, one could formulate it based on longer memory lengths. Taken together, SI captures the difference between the summed entropy of $X$ conditioned on its own past and $Y$ conditioned on its own past (i.e., the two disconnected components) minus the joint entropy of $X$ and $Y$ conditioned on both of their own pasts. Equivalently, SI can be thought of as the average Kullback–Leibler divergence [36] between (i) the joint probability distribution for the present of both $X$ and $Y$ given the past of the connected system, and (ii) the product of the present value probability distributions for $X$ and $Y$ conditioned separately on their own pasts (i.e., 'disconnected' or modeled independently), akin to considerations in Wennekers and Ay [69] and [70].

### 5.1.3 Why:

SI can be applied in computational neuroscience contexts to understand how the activity of two neural processes—from single neurons to ensembles to macroscale regions—evolves through different states together, in a way where the future value of activity in region $X$ depends on its own past and present along with those of region $Y$, and vice versa. A larger value of $\mathrm{SI}(X, Y)$ indicates more shared information between the present and past of the joint distribution of $X$ and $Y$ (i.e., the fully interconnected paired case) compared to information contained in $X$ and $Y$ separately. Since the joint entropy conditioned on the past of $X$ and $Y$ is the subtracted term in Eq. (16) (cf. Fig. 10), a smaller value of $\mathrm{SI}(X, Y)$ suggests that this joint entropy term may be comparatively larger—and therefore that there is not so much information shared in the fully connected paired system.

We depict the SI values between the LOC and all other regions in the left cortex from an example HCP participant in Fig. 4F, computed using the Kozachenko–Leonenko estimator. Here, we observe that the LOC exhibits maximal SI with the caudal anterior cingulate cortex, indicating that these two regions share the most information between their past and present respective BOLD fMRI magnitudes in this example fMRI dataset. By contrast, the LOC shows the lowest SI with the caudal middle frontal cortex, indicating minimal difference in the amount of information contained in these two regions when considered separately rather than jointly. We note that these same regions exhibited corresponding trends with CE (cf. Fig. 4D), consistent with the notion that BOLD magnitude in the LOC maximally reduces uncertainty in the caudal anterior cingulate and minimally reduces uncertainty in the caudal middle frontal cortex in this particular individual. SI has largely been used to quantify the spatiotemporal sharing of information among neural ensembles [71] or sensory processing in the feline auditory cortex [72] and macaque cortex [73]. In a comprehensive functional connectivity benchmarking analysis, Liu et al. [74] reported that SI exhibited among the greatest structure–functional coupling out of hundreds of evaluated measures.

# 6 Pairwise order-dependent measures, directed

## 6.1 Time-lagged mutual information

### 6.1.1 What:

Time-lagged mutual information (TLMI) [75, 76] quantifies the statistical dependence between two time series—specifically, how much information a past (time-lagged) value of one process, $X$, provides about a present or future value of another $(Y)$. TLMI is computed using the mutual information framework (cf. Sec. 2.2), but specifically applied across temporal offsets to capture time-lagged interactions between $X$ and $Y$, without assuming any specific functional form of the relationship.

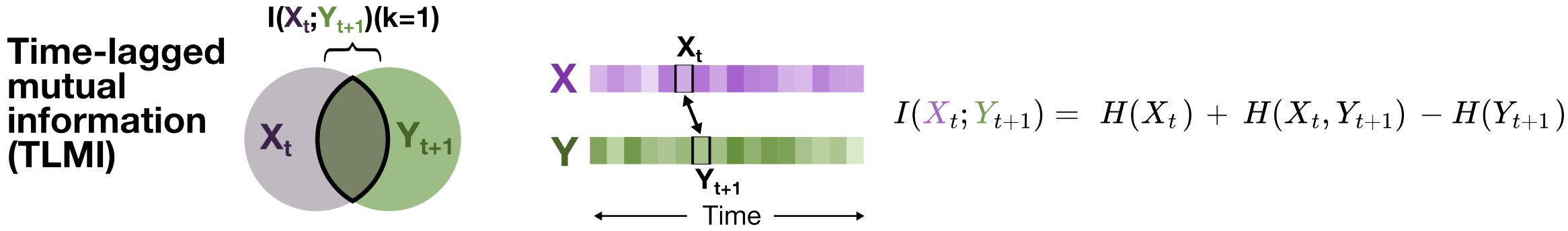

**Figure 11. Time-lagged mutual information.** Time-lagged mutual information (TLMI) quantifies the information shared between the past of one process $(X)$ and the present of another $(Y)$. While MI itself is not a directed measure, the directionality of this measure refers to the sensitivity to which time series ($X$ or $Y$) is lagged. The mathematical form is provided to the right of the schematic diagram and discussed in Sec. 6.1.2.

### 6.1.2 How:

As depicted in Fig. 11, TLMI is computed as

$$
\begin{aligned}
I(X_t;Y_{t+1}) &= H(X_t) + H(Y_{t+1}) - H(X_t, Y_{t+1}), && \textbf{(17)}\\
&= H(X_t) - H(X_t|Y_{t+1}), && \textbf{(18)}\\
&= H(Y_t) - H(Y_{t+1}|X_t), && \textbf{(19)}
\end{aligned}
$$

indicating that $I(X_t;Y_{t+1})$ is equivalent to the difference between the sum of the marginal entropies, $H(X_t)$ and $H(Y_{t+1})$, minus the joint entropy, $H(X_t, Y_{t+1})$ (cf. 'Joint entropy', Sec. 2.1). We classify this as a *directed* pairwise coupling measure due to the asymmetry that arises depending on which of the two time series processes is treated with the lag. In other words, MI itself is not directed, since $I(X_t;Y_{t+1}) = I(Y_{t+1};X_t)$; however, this equivalence does not hold if the lag is reversed, meaning that $I(X_t;Y_{t+1}) \neq I(X_{t+1};Y_t)$.

### 6.1.3 Why:

If $X$ and $Y$ represent the activity in two brain regions (or neuronal ensembles), Greenblatt et al. [77] note that the shift(s) incorporated in TLMI can capture the temporally-lagged influence of activity measured in region $X$ on that in region $Y$ and vice versa. As such, a larger value of $I(X_t;Y_{t+1})$ indicates that the past states of region $X$ share more information with the current state of region $Y$, while a smaller value indicates minimal cross-region dependence. Von Wegner et al. [78] demonstrated that TLMI can better distinguish short- versus long-range memory in neurophysiological time series from both EEG and fMRI compared to the Hurst exponent (which measures variance across timescales [79]). Other applications of TLMI to EEG have yielded connectivity networks with consistent delays across timescales [80] and highlighted the contribution of phase signals to microstate sequences [81, 82].

In our simple empirical example in Fig. 4G, we depict the TLMI values from the past (or lagged) value of the LOC BOLD fMRI time series (realization $x_t$ corresponding to $X_t$ in this example) to the present value of each other target region in the left cortex (realization $y_{t+1}$ corresponding to $Y_{t+1}$ in this example), computed for an example HCP participant using the Kraskov–Stögbauer–Grassberger estimator. This shows that the past of the LOC shares the most information with the current value in the caudal middle frontal cortex. Since we specifically implement this with a time lag of 1 BOLD frame, this suggests that the past state of the LOC is maximally predictive of the next state of the caudal middle frontal cortex in this example fMRI dataset. Of note, the caudal middle frontal cortex exhibited a relatively large MI and the smallest-magnitude CE and JE with the LOC, all of which focus on contemporaneous interactions between the two regions—suggesting that the time-lagged relationship incorporated in TLMI remains relevant to the (potentially nonlinear) contemporaneous interaction between these two regions. Moreover, the maximal TLMI here is a contributing factor to the minimal SI with the caudal middle frontal cortex. By contrast, the LOC exhibits the lowest TLMI with the lateral orbitofrontal cortex as the target, suggesting minimal time-lagged dependence between these two brain regions in this participant.

## 6.2 Causally conditioned entropy

### 6.2.1 What:

Causally conditioned entropy (CCE) [83, 84] is a directed measure, denoted $H(Y\|X)(k)$, that quantifies the uncertainty remaining in one process, $Y$, in the context of its own past up to memory length $k$ ($Y_{t+1}$) and the past *and present* of another process, $X$ (up to memory length $k+1$). In other words, $H(Y\|X)(k)$ captures how much uncertainty remains for the present value of $Y$ when we know its past $k$ values alongside the past and present of $X$ up to time $k+1$. One could set a single value for $k$ *a priori* or evaluate different values up to a maximum window length of $K \ll T$, where $T$ is the length of time series $X$ and $Y$ (see 'Active information storage', Sec. 4.1, for a discussion on choosing $k$ and ergodicity assumptions). Note that the term 'causal' here refers to the conditioning on terms in the past, which *may* be causal to $Y$. However, as an observational rather than interventional measure, CCE does not directly measure causality [85].

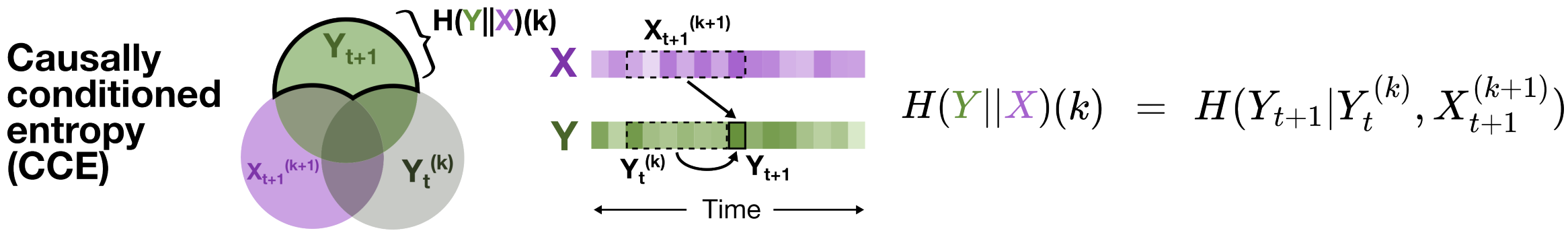

**Figure 12. Causally conditioned entropy.** Causally conditioned entropy (CCE) quantifies the uncertainty remaining in the present of one process ($Y$) after observing its own past along with the past and present of another process ($X$). Note that in this implementation, the memory length $k$ that defines the 'past' is consistent between $X$ and $Y$. The mathematical form is provided to the right of the schematic diagram and discussed in Sec. 6.2.2.

### 6.2.2 How:

As schematically depicted in Fig. 12, the CCE is defined as the entropy of a target process ($Y$) conditioned on both its own past and the past and present of a source process ($X$). More formally, for the stochastic processes $X$ and $Y$, CCE quantifies the uncertainty in the future of $Y$ given its own history and the history of $X$.

Following Equation #5 in Ziebart et al. [84], we define the time-indexed formulation of CCE, $H(Y\|X)(k)$, as Ziebart et al. [84] as

$$H(Y||X)(k) = H(Y_{t+1}|Y_t^{(k)}, X_{t+1}^{(k+1)}), \tag{20}$$

where $Y_t^{(k)}$ denotes the past $k$ values of $Y$, and $X_{t+1}^{(k+1)}$ denotes the past $k$ values plus the present value of $X$ up to time $t+1$. This formulation captures the conditional entropy of $Y$ at the next time point ($t+1$), given its own past *and* the past and present of $X$.

### 6.2.3 Why:

If $X$ and $Y$ represent the activity of two brain regions, a larger $H(Y\|X)$ means that observing past values of both regions does not reduce the uncertainty in the present value of $Y$ by so much—suggesting a weaker dependency between the present value of $Y$ and its own past, along with the past values of $X$. Figure 4H depicts the CCE for all brain regions in the left cortex conditioned on our region of focus, the LOC, in one example HCP participant—computed using the Kozachenko–Leonenko estimator. Here, we observe the maximum CCE in the parahippocampal gyrus, which therefore exhibits the greatest remaining uncertainty in its present value after conditioning on its own past and the past of the LOC in this example fMRI dataset. Our observation that the parahippocampal gyrus exhibits minimal self-predictive information as defined by the AIS (cf. Fig. 4E) would be a key contributing factor to this. We might also expect the parahippocampal gyrus to exhibit relatively low MI and TLMI with the time-lagged LOC, although this is not the case in Fig. 4G. Note that MI and TLMI each measure the influence of only one value of the time series we condition on, while our implementation of CCE (using the *pyspi* library [19, 20]) uses $k = 5$ past values jointly, such that values are not directly comparable anyway. Altogether, the high CCE indicates that the current state of the parahippocampal gyrus depends relatively weakly on its own past and the past BOLD activity in the LOC (when considered jointly) for this participant. The LOC exhibited the lowest CCE with the supramarginal gyrus, which similarly reflects the supramarginal gyrus ranking among the top of all evaluated left cortical regions for AIS values (cf. Fig. 4E)—such that it exhibits high self-predictive information, and the past BOLD activity in the LOC further reduces uncertainty in the current state of the supramarginal gyrus in this individual.

Prior applications in computational neuroscience have predominantly incorporated CCE as a building block for the directed information measure (cf. Sec. 6.3) [86–88]. Through a comprehensive evaluation of over 200 pairwise time-series measures (including several information-theoretic measures) [20], CCE emerged as a top performer in two classification case studies using EEG (to distinguish cortical 'up' or 'down' states, as in Birbaumer et al. [89]) and fMRI (to distinguish resting versus film-viewing, as in Byrge and Kennedy [90]). By contrast, Karimi-Rouzbahani and McGonigal [91] evaluated CCE along with several other directed connectivity measures to characterize neural dynamics in the seizure onset zone during and outside of epileptogenic activity, finding minimal support for predicted patterns with CCE compared to other methods.

## 6.3 Directed information

### 6.3.1 What:

Directed information (DI) is a measure of information flow introduced by Massey [92] to quantify how much information is shared between the present value of one process and the past *and present* of another process. In other words, $\mathrm{DI}(X \to Y)$ captures the total directional influence of $X$ on the present value of $Y$ beyond what can be explained by the past of $Y$ on its own.

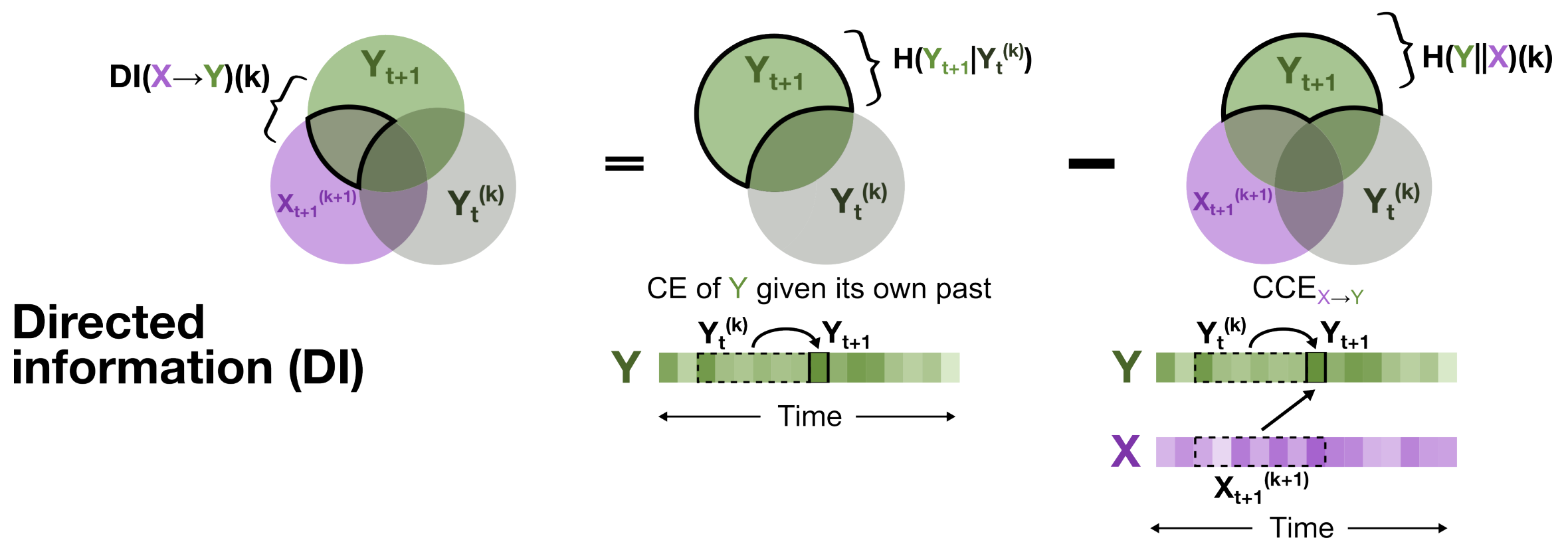

$$DI\,(X \to Y)\,(k) = I(Y_{t+1}; X_{t+1}^{(k+1)} | Y_t^{(k)}) \;=\; H(Y_{t+1}|Y_t^{(k)}) \;-\; H(Y||X)(k)$$

**Figure 13. Directed information.** The sub-component $\mathrm{DI}(X \to Y)(k)$ of directed information (DI) quantifies the information shared between the present of the target time series process ($Y$) and the past and present of another source process ($X$) after conditioning on the probability distribution of $Y$'s past. Note that in this implementation, the memory length ($k$) that defines the 'past' is consistent between $X$ and $Y$. The mathematical form is provided below the schematic diagram and discussed in Sec. 6.3.2.

### 6.3.2 How:

For each of the $T$ samples of the present of $Y$ ($Y_{t+1}, t \in [0, T-1]$, where $T$ is the length of the processes $X$ and $Y$ and the first sample is $t = 1$), DI sums the mutual information from the entire past and present of $X$ (denoted as $X_{t+1}^{(t+1)}$) after conditioning on the entire past of $Y$ ($Y_t^{(t)}$):

$$\mathrm{DI}(X \rightarrow Y) = \sum_{t=0}^{T-1} I(Y_{t+1}; X_{t+1}^{(t+1)} | Y_t^{(t)}), \tag{21}$$

$$= \sum_{t=0}^{T-1} H(Y_{t+1}|Y_t^{(t)}) - H(Y_{t+1}|Y_t^{(t)}, X_{t+1}^{(t+1)}), \tag{22}$$

$$= \sum_{t=0}^{T-1} H(Y_{t+1}|Y_t^{(t)}) - H(Y||X)(t), \tag{23}$$

where Eq. (21) formulates DI in terms of the conditional mutual information between the present of $Y$ ($Y_{t+1}$) and the entire past and present of $X$ ($X_{t+1}^{(t+1)}$), conditioned on the past of $Y$ up to time $t$ ($Y_t^{(t)}$, with $Y_t^{(0)}$ as a null variable by convention). This can be decomposed, as in Eq. (22), into the conditional entropy of $Y$ conditioned on its own past ($H(Y_{t+1}|Y_t^{(t)})$)—which is a form of entropy rate [2, Section 4.2]—minus the conditional entropy of $Y$ conditioned on the past of both $X$ and $Y$ ($H(Y_{t+1}|Y_t^{(t)}, X_{t+1}^{(t+1)})$). The subtracted term in Eq. (22) was introduced in 'Causally conditioned entropy' (cf. Sec. 6.2 [93]) as, equivalently, the CCE with $Y$ conditioned on $X$. The CCE notation, $H(Y\|X)$, is included in Eq. (23), with the addition of $(t)$ to denote the memory length.

As discussed in 'Active information storage' (cf. Sec. 4.1)—and particularly relevant in the bivariate case—estimating this quantity can be empirically challenging due to the high dimensionality of joint past vectors and the fact that only a single realization of the time series exists per time point $t$. To address this, the implementation in *pyspi* [19, 20] imposes a maximum past window length $K \ll T$, where the default is $K = 5$. This constraint allows the algorithm to pool information across the full time series using an ergodic assumption, enabling the computation of a single conditional MI estimate $\mathrm{DI}(X \rightarrow Y)(k)$ for each window length $k$ (up to $K$), rather than estimating it separately for each time point. This is given below in Eq. (24), and carried throughout subsequent equations:[5]:

$$\mathrm{DI}(X \rightarrow Y)(k < K) = \sum_{k=0}^{K-1} DI(X \rightarrow Y)(k), \tag{24}$$

$$= \sum_{k=0}^{K-1} I(Y_{t+1}; X_{t+1}^{(k+1)} | Y_t^{(k)}), \tag{25}$$

$$= \sum_{k=0}^{K-1} H(Y_{t+1}|Y_t^{(k)}) - H(Y_{t+1}|Y_t^{(k)}, X_{t+1}^{(k+1)}), \tag{26}$$

$$= \sum_{k=0}^{K-1} H(Y_{t+1}|Y_t^{(k)}) - H(Y||X)(k). \tag{27}$$

The component $\mathrm{DI}(X \rightarrow Y)(k)$ of $\mathrm{DI}(X \rightarrow Y)$ for a given past window length $k$ is depicted schematically in Fig. 13. Note that the inclusion of the present state of the source ($X$, in this case) is one difference between DI and transfer entropy (discussed in the next section, cf. Sec. 6.4), while another is the sum over time points in DI [94, App. C].

### 6.3.3 Why:

DI provides a perspective of information flow that encompasses both lagged and contemporaneous dependence between the source and target time series. For computational neuroscience, this is particularly useful in mapping how information is stored and propagated across spatial and temporal scales [95]. Malladi et al. [86] used DI to track seizure spread from the epileptogenic zone outwards using electrocorticography (ECoG) recordings in a patient with epilepsy. Other work has demonstrated that DI can model information flows from neural spike train recordings [93, 96], fMRI time series [25], and EEG time series [97, 98]. If $X$ and $Y$ represent the activity of two brain regions, then $\mathrm{DI}(X \to Y)$ captures the information shared between the present of $Y$ and the past *and present* of $X$, after conditioning on the past of $Y$. A larger value of $\mathrm{DI}(X \to Y)$ indicates that knowing the past and present values of activity in region $X$ reduces substantial uncertainty about the present value of region $Y$ beyond that which is reduced by $Y$'s own past. By contrast, a smaller value means that the past and present of activity in region $X$ does not reduce additional uncertainty in $Y$ beyond its own past—which could indicate minimal directed dependence from $X$ to $Y$ (i.e., a larger CCE $X \to Y$), and/or that the past of $Y$ itself maximally reduces uncertainty about the next state of $Y$ (i.e., a smaller CE of $Y$ with its own past)—as schematically depicted in Fig. 13.

[5] The quantity $H(Y_{t+1}|Y_t^{(k)})$ is approximated as $H(Y_{t+1}^{(k+1)})/(k+1)$ in *pyspi* [20], which becomes correct asymptotically as $k \rightarrow \infty$ for stationary stochastic processes [2, Theorem 4.2.1].

In Fig. 4I, we examine the DI from the LOC (source) to all other target regions in the left cortex for BOLD fMRI time series from an example HCP participant, computed with the Kozachenko–Leonenko estimator. This reveals that the maximum DI occurs from the LOC to the inferior parietal cortex, suggesting that the past and present values of the LOC maximally reduce uncertainty remaining in the inferior parietal cortex activity after conditioning on its own past in this example fMRI dataset. In other words, the LOC past and present BOLD activity is most informative of the present inferior parietal cortex activity after accounting for its own prior activity. By contrast, the LOC exhibits minimal DI to the entorhinal cortex, which is consistent with high CCE from the LOC to the entorhinal cortex shown in Fig. 4H—meaning that the LOC does not reduce much of the uncertainty remaining in the entorhinal cortex after conditioning on its own past, indicating minimal directed dependence between the two regions in this individual.

## 6.4 Transfer entropy

### 6.4.1 What:

Transfer entropy is a measure of information flow introduced by Schreiber [75] (see also Bossomaier et al. [99]) to capture the directed dependence from one process ($X$) to another ($Y$) in the context of the past of $Y$. Specifically, TE measures the amount of information shared between the *past values* of $X$ (up to history length $l$) and the present of $Y$ after conditioning on the past $k$ values of $Y$. As mentioned above (see 'Directed information', Sec. 6.3), DI and TE are disambiguated by their inclusion (DI) or exclusion (TE) of the present value of $X$. By not including the present value of the source $X$ in the model for $Y$, TE alone is akin to building a model of how the target process $Y$ is generated or computed from the past of both [94]. Note that, unlike previous order-dependent pairwise features (for which the history length $k$ is consistent between both $X$ and $Y$), TE allows the history length to differ between the two time series. In other words, $\mathrm{TE}(X \to Y)$ captures the total directional influence of the past $l$ values of $X$ on the present value of $Y$ beyond what can be explained by $Y$'s own past (up to length $k$). As with $k$, the memory length $l$ for $X$ can be set to a single value or a range of values can be compared.

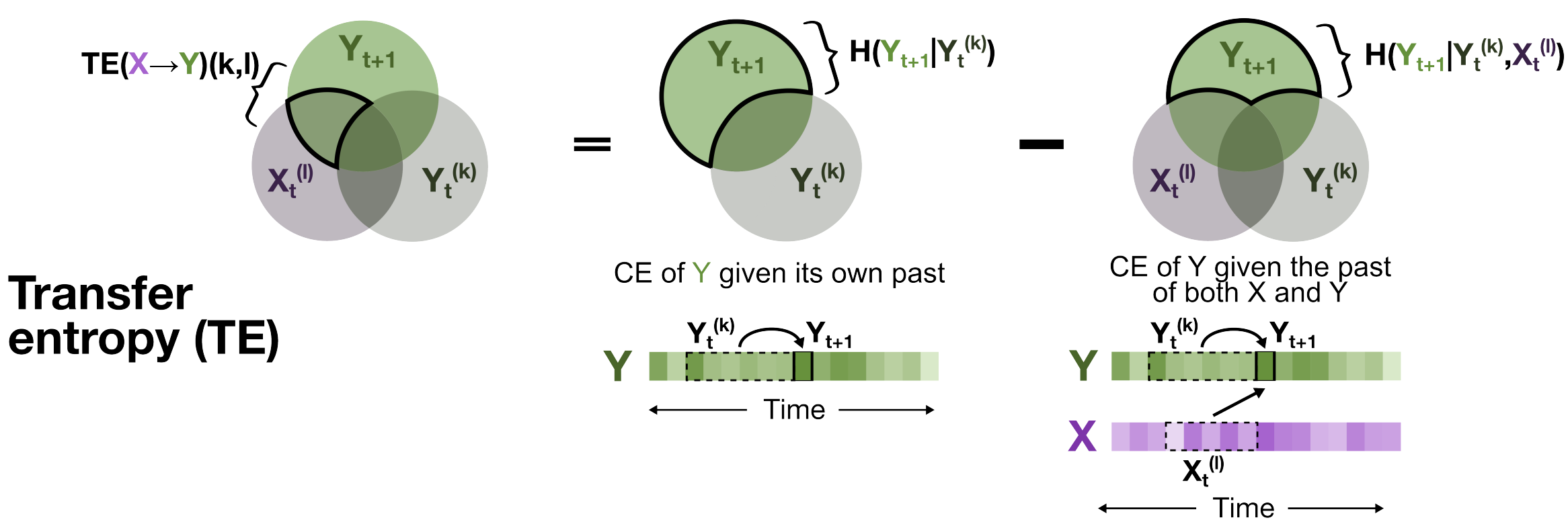

$$TE(X \to Y)(k,l) = I(Y_{t+1}; X_t^{(l)} | Y_t^{(k)}) = H(Y_{t+1}|Y_t^{(k)}) - H(Y_{t+1}|Y_t^{(k)}, X_t^{(l)})$$

**Figure 14. Transfer entropy.** Transfer entropy (TE) quantifies the information shared between the present of one target process ($Y$) and the past of another ($X$) after conditioning on the probability distribution over the past of $Y$. Note that in this implementation, $X$ and $Y$ can be assigned different memory lengths ($l$ and $k$, respectively) that define the past period. The mathematical form is provided below the schematic diagram and discussed in Sec. 6.4.2.

### 6.4.2 How:

As depicted schematically in Fig. 14, TE is given as

$$\mathrm{TE}(X \rightarrow Y)(k,l) = I(Y_{t+1}; X_t^{(l)} | Y_t^{(k)}) \tag{28}$$

$$= H(Y_{t+1}|Y_t^{(k)}) - H(Y_{t+1}|Y_t^{(k)}, X_t^{(l)}), \tag{29}$$

which presents TE as the mutual information (cf. Sec. 2.2) between the present of $Y$ ($Y_{t+1}$) and past of $X$ ($X_t^{(l)}$) after conditioning on the past of $Y$ ($Y_t^{(k)}$) in Eq. (28). Equivalently, as in Eq. (29), TE can be formulated as the CE (cf. Sec. 3.1) of $Y$ given its own past ($H(Y_{t+1}|Y_t^{(k)})$) minus the CE of $Y$ given the past of both itself ($Y_t^{(k)}$) and $X$ ($X_t^{(l)}$). As per the other measures here, TE may be extended for multivariate sources and targets, and may be made conditional on other source variables [25, 100, 101]. Unlike with the DI implementation, the TE implementation in *pyspi* [19, 20] (in utilizing *JIDT* [18]) allows for both fixed default history lengths for $X$ ($l$) and $Y$ ($k$), as well as different delay lag values ($\tau$) for both $X$ and $Y$ (i.e., with the past vector $Y_t^{(k)}$ sampling every $\tau$-th value, from $Y_t$ backwards), as well as flexible optimization based on active information storage values (following [53, 102]). We have assumed $\tau = 1$ for both $X$ and $Y$ in this work for simplicity.

### 6.4.3 Why:

TE has become an increasingly popular measure of directed information flow for neural time series data in the last fifteen years. If $X$ and $Y$ represent activity of two brain regions, a larger $\mathrm{TE}(X \rightarrow Y)$ indicates that knowledge of the past activity of the source region ($X$) reduces more uncertainty about the present value of the target region ($Y$) after already knowing the past of $Y$, suggesting a stronger (potentially nonlinear) influence of the past of $X$ on the present of $Y$. As a measure based on observations, transfer entropy measures contribution to predictability, rather than causal effect [85]. Wibral et al. [39] frame the 'causal' nature often attributed to TE (from this predictive effect) as the following question: Given we already know the past state of region $Y$, how much additional information does the past state of region $X$ provide about the next state of $Y$?

As above, by not considering any contemporaneous dependence between $X$ and $Y$, TE isolates influences of the past values of both $X$ and $Y$ on the future values of $X$ or $Y$. This is in contrast to DI, which also includes contemporaneous dependence between $X$ and $Y$. While TE is a principled choice for measuring how the target is computed or generated from the past of both time series, the time lag has an impact with functional modalities with a lower sampling frequency, such as BOLD fMRI (sampled on the order of seconds, compared to EEG or neural spike recordings with millisecond resolution), as it may be difficult to identify source–target relationships over such a large timescale in comparison to between contemporaneous samples. Given this, TE may be more practically suited to modalities with higher sampling rates, as the dynamic relationship between the past of $X$ and the present of $Y$ may be more fully captured by the data. TE has been applied largely to simulated data from neural mass models [42, 57, 103, 104] and biophysical synapse models and to empirical data with high temporal resolution, like EEG [105, 106], MEG [107] and neural spike trains [108, 109]. However, successful application to BOLD fMRI time series is indeed possible; for example, Lizier et al. [25] applied TE to measure information flows in visuomotor tracking tasks and how they changed with task difficulty, while Taylor et al. [110] applied TE to demonstrate functional attenuation of cholinergic region communication in the macaque brain after locally inactivating the nucleus basalis of Meynert.

In Fig. 4J, we depict TE values from the LOC as the source to all other target regions in the left cortex of an example HCP participant, computed using the Kraskov–Stögbauer–Grassberger estimator with $k = l = 1$. Here, we observe the highest TE from the LOC to the medial orbitofrontal cortex in this example fMRI dataset, suggesting that the past activity in the LOC maximally reduces uncertainty in future activity of the medial orbitofrontal cortex (after conditioning on its own past BOLD fMRI activity). By contrast, the LOC exhibited the lowest overall TE to the superior parietal cortex, indicating minimal time-lagged dependence between these two regions beyond information shared between the past and present activity in the superior parietal cortex itself (which was marked with a relatively high AIS) in this participant.

## 6.5 Granger causality

### 6.5.1 What:

Granger causality (GC) is a measure of information flow introduced by Granger [111] that quantifies the predictive power of the past values of one process $(X)$ adds to an autoregressive model fit to another process $(Y)$. In other words, GC captures how much better a model fit to the past of $Y$ can predict the future of $Y$ when it also has access to the past of $X$. While not necessarily a pure information-theoretic measure by traditional definition, GC is equivalent to TE (see 'Transfer entropy', Sec. 6.4) with a Gaussian density estimator up to a multiplicative factor [112] and is often applied to characterize brain dynamics and connectivity [113], motivating its inclusion here.

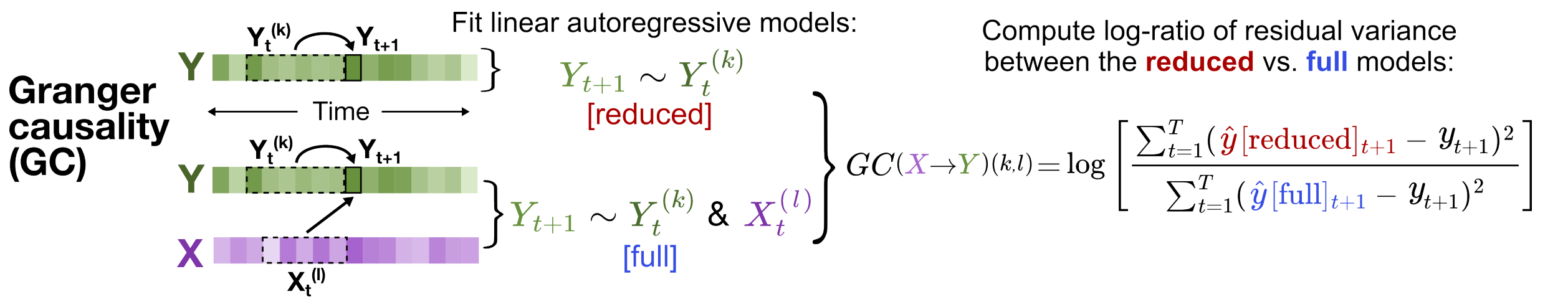

**Figure 15. Granger causality.** Granger causality (GC) quantifies the predictive information that the past of one time series (e.g., $X$) adds to an autoregressive model for the other time series (e.g., $Y$) beyond that of $Y$'s own past. Note that in this implementation, $X$ and $Y$ can be assigned different memory lengths ($l$ and $k$, respectively) that define the past period. The mathematical form is provided to the right of the schematic diagram and discussed in Sec. 6.5.2.

### 6.5.2 How:

As with TE, GC allows for the history length to differ between $X$ (denoted as $l$) and $Y$ (denoted as $k$). Specifically, GC measures the log-ratio of the residual variance for the predicted value of $Y_{t+1}$ based solely on the past values of $Y$ up to window length $k$ ($Y_t^{(k)}$; 'reduced' model) versus that of the predicted value of $Y_{t+1}$ based on the past $l$ values of $X$ ($X_t^{(l)}$) as well the past $k$ values of $Y$ ('full' model). As schematically depicted in Fig. 15, GC is given as

$$\mathrm{GC}(X \to Y)(k,l) = \log \left[ \frac{\sum_{t=1}^{T} (\hat{y}[\text{reduced}]_{t+1} - y_{t+1})^2}{\sum_{t=1}^{T} (\hat{y}[\text{full}]_{t+1} - y_{t+1})^2} \right], \tag{30}$$

where $\hat{y}$ is the vector of values that the autoregressive model predicted for the realization $y$ at each time point $t+1$, in which $\hat{y}[\text{reduced}]$ corresponds to the 'reduced' model (i.e., only the past of $y$) and $\hat{y}[\text{full}]$ corresponds to the 'full' model (i.e., the past of both realizations $y$ and $x$). The numerator in the logarithmic fraction in Eq. (30) corresponds to the sum of squared differences between each $\hat{y}_{t+1}$ (predicted) minus $y_{t+1}$ (real) value—that is, the squared residuals of the 'reduced' model. Similarly, the denominator contains the sum of the squared residuals from the 'full' model.

In *pyspi* [19, 20], GC is computed from the same code as TE with the Gaussian density estimator due to their equivalence as above.[6] For an in-depth discussion about the similarities between GC and TE, we direct the interested reader to Milinkovic et al. [114]. Of note, Granger causality can also be computed in the frequency domain, as spectral Granger causality [115]. This allows the contribution of different frequencies to the dependence to be inferred.

### 6.5.3 Why:

GC is very widely used as an index of 'causal' interactions between brain areas in the computational neuroscience literature [113, 116, 117]. Like transfer entropy, 'causal' here is with reference to a *predictive* model rather than considering interventional notions. If $X$ and $Y$ represent the activity of two brain regions, then $\mathrm{GC}(X \to Y)$ captures the additional predictive information that the past of region $X$ provides for the future of region $Y$ beyond the past activity of $Y$ alone. Multivariate extensions have also been developed in the pursuit of 'full-brain connectivity' representations [118, 119].

[6] In principle, Granger causality is a unitless measure; however, given the known equivalence to TE above, it is often expressed in nats when computed in that way.

In Fig. 4K, we examine the GC values from the LOC as the source to all other (target) regions in the left cortex (using $k = l = 1$), noting the maximum value detected in the fusiform gyrus. This indicates that incorporating information about the past activity of the LOC contributes more predictive information to the future of the fusiform gyrus than an autoregressive model fit to the past activity of the fusiform gyrus alone. Of note, the fusiform gyrus exhibited the maximal entropy of all evaluated left cortical regions in this participant (cf. Fig. 4A), suggesting that activity in this region is more unpredictable over time—which provides plenty of uncertainty for the past activity of the LOC to potentially reduce when it is included. We also note that Granger causality assumes a joint Gaussian bivariate distribution, unlike the pure (conventional) information-theoretic measures covered here, especially the TE, which do not inherently assume the form of underlying data (see 'A note on estimating MI in continuous space', Sec. 2.2). By contrast, we observe the lowest GC from the LOC to the lingual gyrus, indicating that the past activity in the LOC does not substantially reduce uncertainty in the future of the lingual gyrus beyond its own past activity. This is particularly interesting because the LOC and lingual gyrus exhibited the maximal MI in Fig. 4C, which indicates that this inter-regional relationship is stronger with a contemporaneous measure that does not condition out the past of the target, and does not assume a joint bivariate Gaussian distribution. Moreover, we note the contrast in results between GC in Fig. 4K and TE in Fig. 4J—with the two measures only differing in their estimators, this contrast reflects the difference observed in information flow when only modeling linear interactions (GC) versus non-linear (TE).

# Conclusions and future directions

Granger Causality marks the final measure covered in this review, and thus the conclusion of the in-depth coverage of the eleven information-theoretic time-series measures. Collectively, we have presented a unified and accessible guide to these eleven key measures for analyzing time series data, with a particular focus on their applications in computational neuroscience. Through a novel schematic overview (with measure-specific decompositions in each section) and a systematic tabular summary, we have sought to clarify conceptual distinctions between measures—such as directionality, temporal dependence, and dimensional scope—and to bridge terminology across disparate literatures.

Our aim has been not only to consolidate fundamental definitions but also to offer an integrative perspective that enables researchers with a wide range of background knowledge depths to navigate this methodological landscape with greater fluency. While our examples and emphasis here have focused on neural time-series data, we note that all measures are broadly applicable across diverse scientific domains. Resources like the UEA/UCR Time Series Classification repository [120, 121] provide compelling examples of diverse time-series datasets—from beef spectrograms to yoga pose angles to goose vocalizations—that could benefit from this conceptual framework to yield new insights into how information-theoretic measures relate to each other when computed on diverse empirical data. Each measure in this guide offers a unique lens on dynamics such as uncertainty, dependence, and predictability. Developing an intuitive grasp of how these measures interrelate and differ lays the foundation for richer, more integrative analyses of complex systems like the brain. We hope this guide serves as a starting point, upon which future work may extend this foundation by incorporating additional measures, refining estimation techniques, or exploring emerging applications.

# Methods

## Neuroimaging preprocessing for case study

For the empirical neuroimaging case study, we focused on resting-state fMRI (rs-fMRI) data from one randomly selected participant (ID '298051') from the S1200 release of the Human Connectome Project [21] (specifically, this participant is included in the subset of 100 unrelated individuals). All resting-state volumes in this dataset were acquired using a gradient-echo, echo planar image (EPI) sequence with the following parameters: TR/TE = 720/33.1 ms, slice thickness = 2.0 mm, 72 slices, 2.0 mm isotropic voxels, frames per run = 1,200. Of note, all imaging data included in the present work were already preprocessed in previous work [27, 122], with a brief overview provided as follows. The rs-fMRI time series were linearly detrended, and the mean signal from white matter, cerebrospinal fluid, and global signal was regressed out. Low-frequency fluctuations ($f < 8 \times 10^{-3}$ Hz) were removed using a high-pass filter, applied to the EPI data using a fast Fourier transform. The average voxelwise time series was summarized per parcel using the Desikan-Killiany 68-region atlas [123], noting that we focused on the left hemisphere in this work (i.e., 34 cortical regions) for simplicity. Information-theoretic measures were computed from the time series using the *JIDT* [18] library, either directly (for the single-process measures) or embedded within the *pyspi* [20] toolbox, both using Python. We applied continuous estimators to the time series, opting for the Kozachenko-Leonenko (`kozachenko`) estimator for measures that are constructed from the univariate or joint entropy—which uses a nearest-neighbor approach—and usually the Kraskov–Stögbauer–Grassberger (`kraskov`) estimator for the other measures, which combines estimates based on mutual information with 4 nearest neighbors. The implemented estimator is specified for each measure.

## Advanced concepts about estimation

In this final section, we address three nuanced yet important considerations when applying information-theoretic measures to empirical data: ergodicity, autocorrelation, and uncertainty estimation. These are more advanced concepts that touch on important potential pitfalls for empirical applications; specifically, we discuss potential violations of notable assumptions in a given dataset and how a practitioner might detect and correct for these.

In this work, we use 'ergodic' to mean that the given realization ($\mathtt{x}$) will reproduce the probabilities of the process space ($\mathtt{X}$), effectively applying the Boltzmann ergodic hypothesis: that long-time averages converge to become equivalent with ensemble averages in an ergodic system [14]. For a finite sample, this may require very long time series to properly capture the probabilistic properties of the process—a requirement that is not always satisfied by short empirical datasets, such as BOLD fMRI neuroimaging. Indeed, mounting evidence indicates that fMRI time series may not satisfy ergodicity even asymptotically, such that longer sampling does not guarantee convergence of certain information-theoretic estimates [124, 125]. This means absolute estimates of information-theoretic measures may be inflated or unstable in these circumstances; importantly, relative comparisons can still be informative in such cases, as in the illustrative functional neuroimaging case study in this work. More generally, when computing a given statistic on a given realization, one must exercise caution about the extent to which this inference can meaningfully translate back to the underlying process.

We do note that the ergodicity assumption, as we introduced here, is generally implicitly baked into information-theoretic estimation from single time-series realizations to enable tractable computation. Whether this assumption is meaningfully approximated by a given empirical dataset—particularly short, autocorrelated fMRI time series—is a separate, and indeed nontrivial, question. Importantly, there is currently no domain-general diagnostic for ergodic convergence of information-theoretic estimators for short, autocorrelated neuroimaging time series, and any such assessment would necessarily be estimator- and preprocessing-specific. Example diagnostics have been developed in statistical mechanics based on statistical symmetry [126] and dynamical systems [127]. Alternatively, for functional neuroimaging applications, one might quantify the stability of information-theoretic measure estimates as a function of imaging window length or examine split-half reliability across runs, analogous to the $\hat{R}$ diagnostic statistic for convergence in Markov chain Monte Carlo methods [128]. As these collectively comprise advanced concepts beyond the scope of this work, we instead refer the interested reader to relevant reading about rigorous ergodicity checks (particularly in the case of relatively short time series) [129–131].

A related consideration is that of stationarity and autocorrelation; specifically, our illustrative case study employed BOLD fMRI, a modality known to be nonstationary and poorly sampled in time. While its statistical properties change over time, BOLD fMRI also exhibits inherent autocorrelation due to neurovascular and physiological processes (namely, the hemodynamic response) that give rise to strong short-range temporal dependencies [132]. Moreover, digital filtering introduced in some preprocessing pipelines to reduce non-stationarity induces an impulse response that can further compound autocorrelation [133]. This raises two key issues when applying information-theoretic measures to BOLD fMRI: (1) the autocorrelation violates the general assumption of independence of samples; and (2) the number of samples is too short for a nonstationary phenomenon. On the first point, it is known that autocorrelation (in BOLD fMRI) affects estimates of both Granger causality and mutual information, for example, by inflating bias when no relationship exists between the variables [133]. This work examined the same broader Human Connectome Project (HCP) resting-state fMRI dataset, comprised of 1200 timepoints per scan—characterized by approximately 40-50 dominant wavelengths. While the details are beyond the scope of the present work, previous work from our group delivered approaches to address autocorrelation for linear estimators [133], paralleling the use of Theiler or dynamic correlation exclusion windows [134] for the nearest-neighbor based estimators (i.e., Kozachenko–Leonenko and Kraskov–Stögbauer–Grassberger)

In terms of non-stationarity, certain fMRI preprocessing steps can help stationarize the time-series data, partly mitigating some issues. We analyzed a single subject selected at random from the broader HCP dataset, preprocessed previously [27] to remove low-frequency components through high-pass filtering at $8 \times 10^{-3}$ Hz, which yields a more stationary signal such that drift is less of a concern. For future work applying information-theoretic measures to a functional neuroimaging dataset at scale (i.e., for quantitative analysis across multiple individuals), additional preprocessing steps can reduce outlier sensitivity and autocorrelation. For example, rank ordering (akin to a Spearman correlation) [37, 124] and pre-whitening [133, 135] can render computed measures less sensitive to outliers. Regressing out autocorrelation in a preprocessing pipeline can also improve the effective sample size and mitigate some practical consequences of slow convergence. We emphasize that these considerations are provided to contextualize potential interpretive pitfalls rather than to prescribe a standardized neuroimaging preprocessing workflow.

Finally, we turn to uncertainty estimation in the context of statistical inference for information-theoretic measures. Barnett and Bossomaier [136] derived the uncertainty of transfer entropy and related (conditional) mutual information measures as $\chi^2$ distributions for certain estimators, providing an analytic parametric inference method. This is a mathematically elegant approach, but it makes strong assumptions about stationarity that often do not hold with neurophysiological data (e.g., BOLD fMRI). One potential solution is to derive modified hypothesis tests with reduced degrees of freedom (i.e., 'effective' degrees of freedom) for applications to correlation-based measures of linear coupling between (covariance-stationary) time series [133]. Additionally, acknowledging that group-level inference does require explicit statistical modeling of uncertainty, Combrisson et al. [137] suggest several options for non-parametric uncertainty quantification, including group-level surrogate data testing combined with individual-level bootstrap resampling. Such approaches construct an empirical null distribution without imposing assumptions about the estimator's sampling distribution. As a detailed discussion of parametric versus non-parametric uncertainty estimation is beyond the purview of this work, we refer the interested reader to both Barnett and Bossomaier [136] and Combrisson et al. [137].

## Data and code availability

All HCP data, including the resting-state fMRI data used in this work, is freely available for registered researchers at https://www.humanconnectome.org/ as described in Van Essen et al. [21]. All code used for empirical analysis and visualization in this work is freely available at https://github.com/anniegbryant/info_theory_visuals, which includes parcellated fMRI time series data for one HCP participant.

## Acknowledgments

We thank Chanelle Noble and Florian Burger for insightful discussions that helped shape this work.

## Funding statement

A.G.B. was supported by the Research Training Program scholarship and stipend (Australian Government) and the Paulette Isabel Jones Career Award (The University of Sydney). J.M.S. was supported through the Australian Research Council (ARC) Discovery Project grant (DP240101295) and National Health and Medical Research Council (GNT2033162). B.D.F. was supported through an ARC Future Fellowship grant (FT240100418) and the ARC Discovery Project grant (DP240101295). J.T.L. was supported through the ARC Discovery Project grant (DP240101295).

# Supplementary materials

**Table S1. Summary of empirical computation times for information-theoretic measures.** The median computation time (seconds) $\pm$ inter-quartile range is summarized for each information-theoretic measure and estimator. Values are averaged across left-hemisphere cortical regions for single-process measure or across region–region pairs (with lateral occipital cortex as the source) for pairwise measures. Blank entries indicate measure–estimator combinations that are not defined or not implemented (for example, Kraskov–Stögbauer–Grassberger for absolute entropy-based measures like DI). Computation times span several orders of magnitude, with estimator choice accounting for larger differences in runtime than the choice of the measure itself. Of note, $k$-nearest-neighbor-based ($k$-NN-based) estimators incur substantially larger time costs for measures involving high-dimensional joint or conditional densities (e.g., SI and DI), while Gaussian estimators provide uniformly fast computation under linear–Gaussian assumptions.

| Measure | Gaussian | Kozachenko-Leonenko | KSG |
|---|---|---|---|
| ENT | 4e-06 $\pm$ 2e-06 | 4e-02 $\pm$ 4e-03 | |
| AIS | 4e-06 $\pm$ 1e-06 | | 2e-03 $\pm$ 3e-04 |
| CCE | 6e-06 $\pm$ 1e-06 | 7e-06 $\pm$ 1e-06 | |
| CE | 3e-04 $\pm$ 8e-05 | 4e-02 $\pm$ 3e-02 | |
| TLMI | 3e-04 $\pm$ 2e-04 | | 2e-03 $\pm$ 7e-04 |
| MI | 4e-04 $\pm$ 3e-04 | | 2e-03 $\pm$ 6e-04 |
| GC | 6e-04 $\pm$ 2e-04 | | |
| JE | 1e-03 $\pm$ 2e-04 | 4e-02 $\pm$ 3e-02 | |
| TE | | | 3e-03 $\pm$ 1e-03 |
| SI | 7e-03 $\pm$ 4e-04 | 2e-01 $\pm$ 1e-01 | |
| DI | 4e-02 $\pm$ 3e-02 | 6e-01 $\pm$ 1e-01 | |

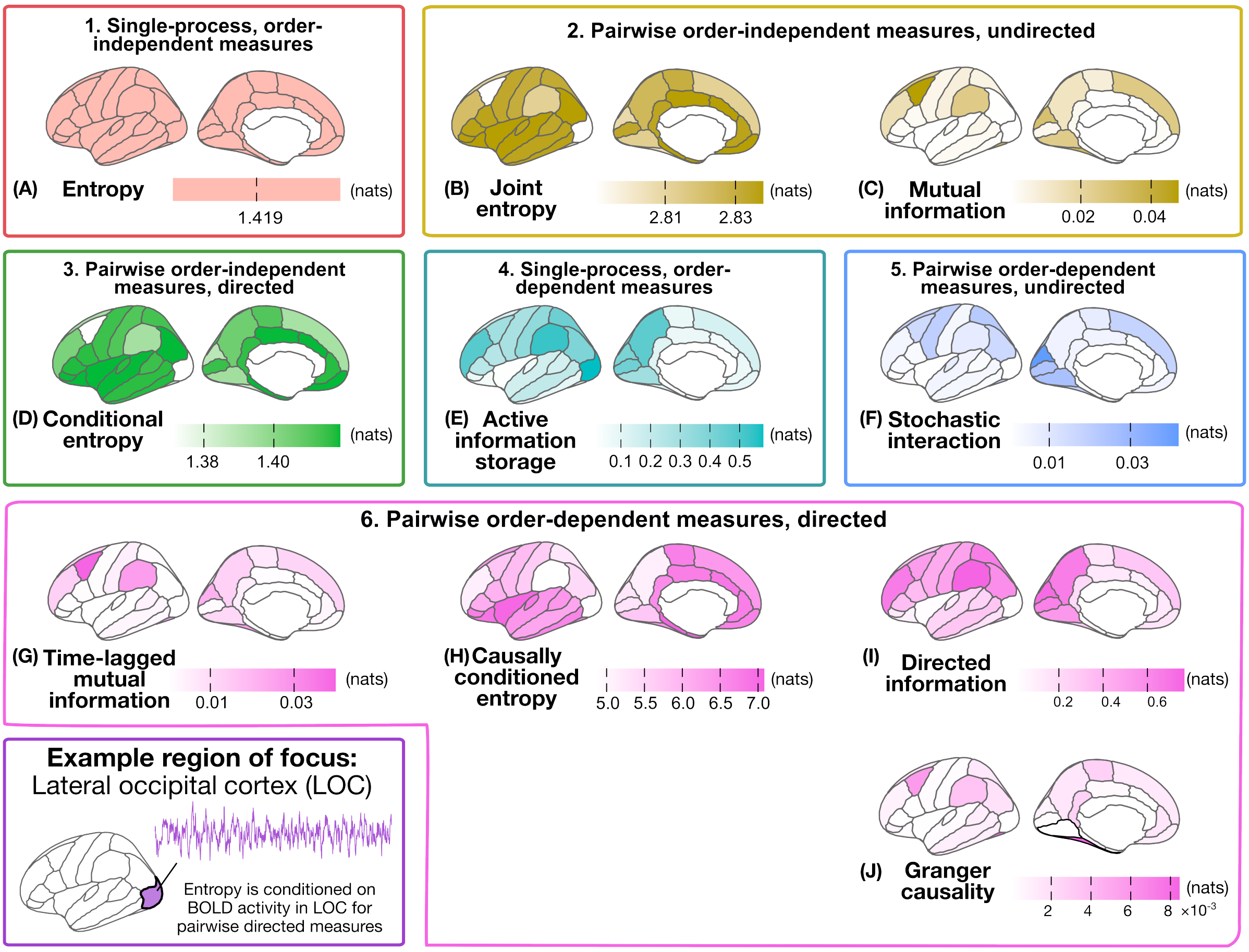

**Figure S1. Visualizing information-theoretic time-series measures on the cortical surface with Gaussian density estimators.** Information-theoretic measures are projected onto the left hemisphere cortical surface as in Fig. 4, here computed using the Gaussian density estimator. In the case of the two single-process measures, entropy and active information storage, the values are computed and shown for each individual region. As Granger causality (GC) and transfer entropy (TE) using the Gaussian density estimator have a one-to-one correspondence, the brain map is shown solely for GC here. For the eight pairwise measures in total, we set the lateral occipital cortex (LOC) as the seed region (shown in purple at the bottom left) and computed the pairwise measures to each of the other 33 regions in the left cortex; the values shown for each brain region reflect the value from the lateral occipital cortex to the corresponding region for the indicated measure.

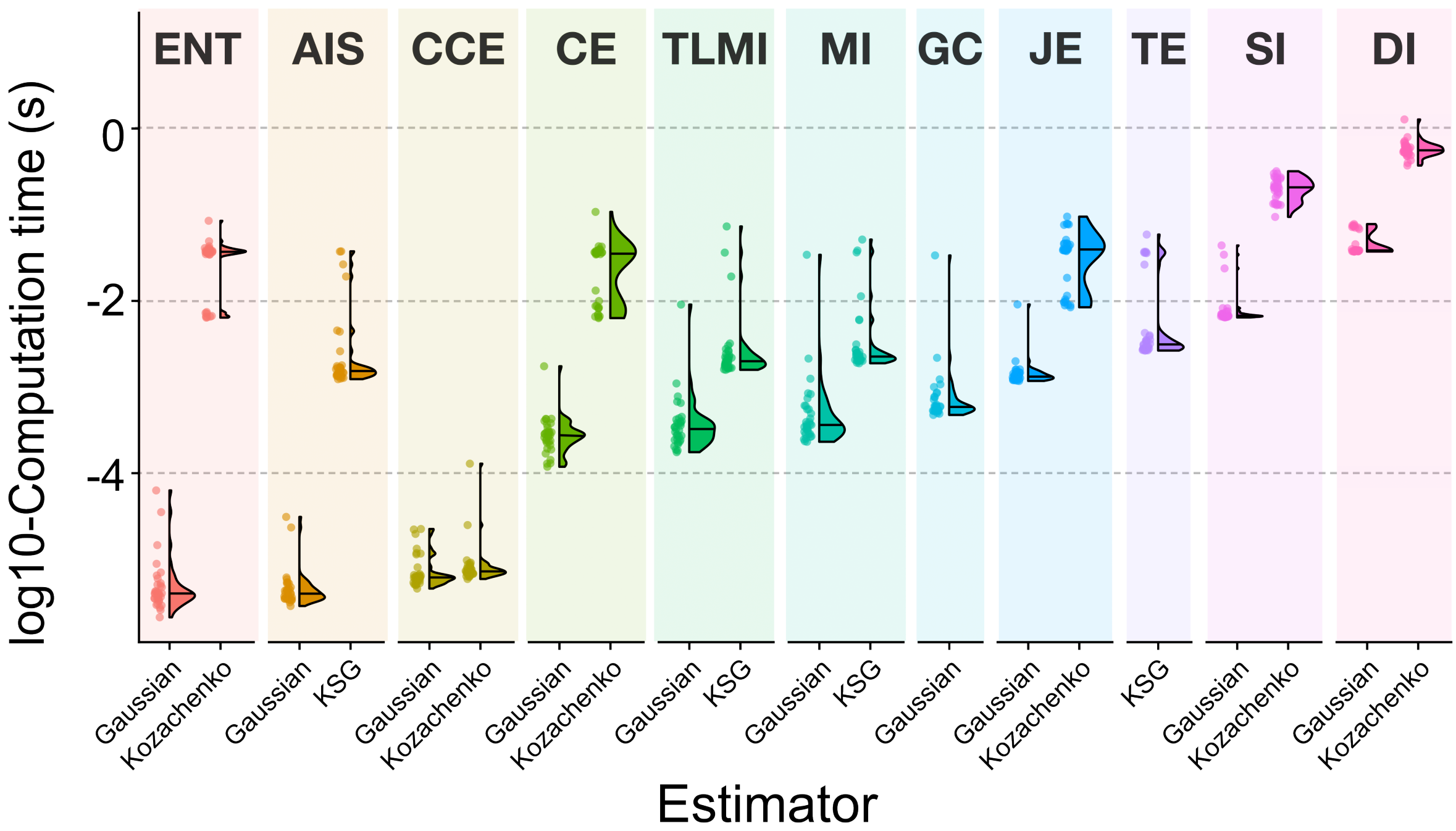

**Figure S2. Computation time for each information-theoretic measure applied to resting-state fMRI data in the illustrative case study.** Raincloud plots showing the distribution of $\log 10$-transformed computation times (in seconds) for each information-theoretic measure computed for an example participant (ID 298051) in the HCP dataset. Pairwise measures were computed from a fixed source region (lateral occipital cortex) to each of the remaining 33 regions in the left cerebral cortex of the Desikan-Killiany parcellation [123]. For single-process measures (ENT and AIS), computation times are shown across all 34 left-hemisphere cortical regions. Each measure was evaluated using one or more continuous estimators (Gaussian, Kozachenko–Leonenko, and/or Kraskov–Stögbauer–Grassberger) where applicable. Half-violins show the empirical distribution across region–region pairs (pairwise measures) or regions (single-process measures), with horizontal lines indicating the median computation time. Across measures, the Gaussian estimators exhibited substantially lower computation times than the $k$-nearest-neighbor-based ($k$-NN-based) estimators, reflecting their closed-form parametric formulation and lack of neighbor searches. Directed and pairwise measures generally show the highest computation times, especially under $k$-NN-based estimation, consistent with the increased dimensionality of the underlying conditional density estimates.